\documentclass[12pt,letterpaper]{amsart}

\usepackage{amsmath,amssymb,amsthm,mathtools}
\usepackage{graphicx,booktabs,array}
\usepackage{float}
\usepackage[round]{natbib}
\setcitestyle{aysep={}}
\usepackage{enumitem}
\usepackage[margin=1in]{geometry}
\usepackage{setspace}
\usepackage{microtype}
\usepackage{needspace}
\usepackage{placeins}
\usepackage{xurl}
\usepackage{hyperref}

\hypersetup{
  colorlinks=true,
  linkcolor=blue,
  citecolor=blue,
  urlcolor=blue,
  pdftitle={Who Picks First? Compensatory Fairness When Equal Division Is Impossible},
  pdfauthor={Aleksei Chernulich and M. Bumin Yenmez},
  pdfkeywords={fair division, compensation, indivisible goods, snake draft}
}

\newtheorem{theorem}{Theorem}
\newtheorem{proposition}{Proposition}
\newtheorem{lemma}{Lemma}
\newtheorem{corollary}{Corollary}
\theoremstyle{definition}

\newtheorem{example}{Example}
\theoremstyle{remark}
\newtheorem{remark}{Remark}

\makeatletter

\let\WPF@hyper@makecurrent\hyper@makecurrent
\def\hyper@makecurrent#1{%
  \def\WPF@tempa{#1}\def\WPF@tempb{subsection}%
  \ifx\WPF@tempa\WPF@tempb\ifx\Hy@chapapp\Hy@appendixstring
      \WPF@hyper@makecurrent{appsubsection}%
    \else\WPF@hyper@makecurrent{#1}\fi
  \else\WPF@hyper@makecurrent{#1}\fi}
\makeatother

\newcommand{\cP}{\mathcal{P}}
\newcommand{\cW}{\mathcal{W}}
\newcommand{\cA}{\mathcal{A}}
\newcommand{\PD}{\mathrm{PD}}
\newcommand{\dom}[1]{\succsim^{\PD}_{#1}}
\newcommand{\sdom}[1]{\succ^{\PD}_{#1}}
\newcommand{\Dr}{D}
\newcommand{\NW}{\textsf{NW}}
\newcommand{\EFO}{\textsf{EF1}}
\newcommand{\NEL}{\textsf{NEL}}
\newcommand{\AAU}{\textsf{AAU}}
\newcommand{\NEU}{\textsf{NEU}}
\newcommand{\MM}{\textsf{MM}}
\newcommand{\IUC}{\textsf{IUC}}
\newcommand{\ORa}{\textsf{OR}}
\newcommand{\EFF}{\textsf{EFF}}
\newcommand{\NB}{\textsf{NB}}
\newcommand{\WSP}{\textsf{WSP}}
\newcommand{\SP}{\textsf{SP}}

\title[Who Picks First?]{Who Picks First? Compensatory Fairness When Equal Division Is Impossible}
\thanks{Chernulich: Department of Economics, Durham University, \href{mailto:aleksei.chernulich@durham.ac.uk}{aleksei.chernulich@durham.ac.uk}. Yenmez: Department of Economics, Washington University in St.\ Louis; Durham University; and \"Ozye\u{g}in University, \href{mailto:bumin@wustl.edu}{bumin@wustl.edu}. We used Claude and ChatGPT to check the writing and the proofs. The authors take full responsibility for the content of the paper.}

\author[Chernulich]{Aleksei Chernulich}
\author[Yenmez]{M. Bumin Yenmez}

\date{\today}

\begin{document}

\begin{abstract}
When indivisible objects, such as heirlooms or tickets to sports events, do not divide evenly among agents with equal claims, some agents must receive more objects than others. Can picking first compensate those with fewer objects? We propose two principles of compensatory fairness. An agent with fewer objects must not envy a larger bundle with any one of its objects removed, and priority advantages among agents with equally many objects must alternate rather than accumulate. Non-wastefulness, neutrality and an object-wise form of Maskin monotonicity characterize the draft rules, in which agents pick in a fixed sequence. Adding the two principles and envy-freeness up to one object singles out the block--snake draft, in which agents with fewer objects pick first in every round and each group's order reverses between rounds. Among draft rules, compensation is compatible with strategy-proofness, but envy-freeness up to one object is not: once at least two agents receive more than one object, no draft that satisfies it is even weakly strategy-proof, and with two such agents the conflict extends to all non-wasteful and neutral rules. When agents may be indifferent between objects, resolving each tie in favor of the earliest later picker who strictly cares preserves the draft's fairness guarantees and ensures efficiency under pairwise dominance.


\end{abstract}

\maketitle

\section{Introduction}\label{sec:intro}

\smallskip
\begin{flushright}\small
\begin{minipage}{0.5\textwidth}\raggedright
\emph{So the last shall be first, and the first last.}\par\smallskip
\hfill--- Matthew 20:16 (King James Version)
\end{minipage}
\end{flushright}
\medskip

When people with equal claims divide objects that cannot be split, the division rarely comes out even. Heirs dividing a parent's paintings or books,\footnote{\citet{PrattZeckhauser1990} study a closely related setting, the actual division of silver heirlooms among eight grandchildren without direct monetary payments.} housemates splitting the furniture they bought together, and friends sharing a set of season tickets all face a remainder whenever the number of objects is not a multiple of the number of people. A Premier League club, for instance, plays $19$ home league games, so a group of two, three or four fans who jointly hold a season ticket cannot split them evenly. Even when bundle sizes differ by at most one, some people receive one object more than others, which can give those who receive fewer grounds for complaint. 

Without money, indivisible objects are often distributed between agents with equal claims by allowing them to take turns.\footnote{We leave aside money, repetition and scheduling constraints. With money, the extra objects could be priced or auctioned, but many groups avoid side payments. With repeated seasons, the extra objects could rotate, but the problem remains within a season and whenever the group changes. When bundles must respect schedules, as with seats in courses or shifts in a roster, feasibility constraints become central.} In fact, alternation is among the simplest procedures for dividing property in divorces and estates \citep{BramsTaylor1999}, and Harvard Business School allocated elective courses by a draft whose order reversed every round \citep{BudishCantillon2012}. When everyone receives the same number of objects, this snake format is common, but when the objects do not divide evenly, it leaves the extra objects to whoever happens to pick in the last round. For the cases where claims are unequal, priority usually rises with the share, as in the picking sequences studied for unequal entitlements \citep{CSS2021}. One service that runs ticket drafts, for instance, gives those who buy more games a greater chance at the top picks.\footnote{See the draft-order algorithm of SplitSeasonTickets:\newline\url{https://public.splitseasontickets.com/home/draftOrder}.} There the unequal numbers are entitlements, paid for in advance. With equal claims the extra objects are no one's due, which suggests that those who receive fewer should perhaps be compensated.

When money cannot change hands and some inequality is unavoidable, the only available compensation may be priority, that is, the right to choose first. Normative economics has long asked how to compensate people for inequalities they are not responsible for \citep{FleurbaeyManiquet2011}. Here the inequality is created by arithmetic, and proposed compensation takes the form of earlier turns. 
What does fairness require here?\footnote{The question is especially relevant when each person receives only a few objects, so that one additional object represents a substantial fraction of a bundle. \citet{BudishCheKojimaMilgrom2013} make a related observation: their bound on variation in an agent's utility across realized allocations, expressed in terms of a single-unit utility difference, is most powerful when agents demand many goods each. The relevant distinction is the number of objects per person, rather than the number of participants or objects alone.} A fair rule must decide who receives the extra objects and how those who receive fewer are compensated. Picking first is a natural form of compensation, since it lets a person choose the objects she wants most. But priority is itself a source of inequality. Among people with the same number of objects, an early pick is an advantage, and if the same person always picks first, the advantage compounds. Fair priority therefore has two sides: compensation for receiving fewer objects, and a limit on how far priority can accumulate among people who receive the same number.

Consider the following illustrative example. Three fans, Alice, Bob and Carol, share tickets to seven games. A lottery orders them Alice, Bob, Carol, and the fan drawn last, Carol, gets the extra ticket, so she receives three tickets and the others two each. All three rank the games identically, and we number them from $1$, the most preferred, to $7$, the least preferred. Compensation lets Alice and Bob pick before Carol in each of the first two rounds, and non-accumulation reverses their order in the second round. Picking in the order Alice, Bob, Carol in the first round, Bob, Alice, Carol in the second, and Carol alone in the third gives Alice $\{1,5\}$, Bob $\{2,4\}$ and Carol $\{3,6,7\}$. Alice's head start in the first round is returned to Bob in the second. Alice and Bob's earlier picks compensate them for receiving one ticket fewer. They do not envy Carol's bundle after any one ticket is removed, since both prefer their own tickets, game by game, even to Carol's two best tickets, $3$ and $6$. We call this a \emph{block--snake} order. In this draft rule, fans with fewer tickets pick first in every round, and among fans with the same number of tickets the order reverses from round to round.

\begin{table}[t]
\centering\small
\begin{tabular}{llccccccc}
\toprule
Rule & Picking order & Alice & Bob & Carol && \EFO & \NEL & \AAU\\
\midrule
Round-robin & A B C $\mid$ A B C $\mid$ A & $\{1,4,7\}$ & $\{2,5\}$ & $\{3,6\}$ && yes & no & no\\
Plain snake & A B C $\mid$ C B A $\mid$ A & $\{1,6,7\}$ & $\{2,5\}$ & $\{3,4\}$ && yes & no & yes\\
Block--snake & A B C $\mid$ B A C $\mid$ C & $\{1,5\}$ & $\{2,4\}$ & $\{3,6,7\}$ && yes & yes & yes\\
\bottomrule
\end{tabular}
\caption{Three fans share seven tickets, numbered from $1$, the most preferred by all three, to $7$, the least preferred. A, B and C stand for Alice, Bob and Carol, and bars separate rounds. All three rules use the order drawn by the lottery, Alice, Bob, Carol. Round-robin repeats the same order every round, the plain snake reverses it, and the block--snake puts the fans with fewer tickets first in every round and reverses the order within each group. The last three columns record whether each rule satisfies the desirable conditions defined in the text. Note that only the block--snake rule satisfies all three.}\label{tab:intro}
\end{table}

\autoref{tab:intro} compares the block--snake draft with round-robin and the plain snake. With the same draw, round-robin and the plain snake give the extra ticket to Alice, who also picks first. In each, Bob's tickets are not at least as good, game by game, as Alice's with her least preferred ticket removed, and under round-robin Bob is also ahead of Carol in both rounds.

We state compensation and limits on priority as principles of compensatory fairness, which are conditions on who receives what. From here on we call the fans \emph{agents} and the tickets \emph{objects}. The first condition, \emph{no envy toward larger bundles, up to any object} (\NEL), requires that an agent with fewer objects not envy a larger bundle, whichever single object is removed from it. This applies envy-freeness up to any good \citep{Caragiannis2019} in one direction only, from smaller bundles toward larger ones. The second, \emph{alternating advantage under unanimity} (\AAU), limits how far priority can accumulate. When all agents rank the objects identically, so that the allocation reflects nothing but priority, compare two agents with the same number of objects by their most preferred objects, then by their second most preferred, and so on. Whoever is ahead at one step must be behind at the next. A first-move advantage may compensate a quantity disadvantage, but it must not become a persistent privilege among equals. For all pairs of agents we also use \emph{envy-freeness up to one object} (\EFO), under which any envy must disappear once a suitable single object is removed from the envied bundle \citep{Budish2011}. We use its ordinal, pairwise-dominance version: the comparison holds for every positive additive utility consistent with the ranking. This is stronger than \EFO\ evaluated at a single utility profile.\footnote{Bundle sizes are not part of the model. Since a bundle cannot dominate a larger one, \EFO\ keeps any two bundles within one object of each other, so when every object is allocated, some agents receive one more object than the others as a consequence of the axioms, not as an assumption (\autoref{lem:sizes} in \autoref{app:aux}).} \NEL\ strengthens \EFO\ when the envied bundle is larger, since there the envy must disappear whichever object is removed. As the last three columns of \autoref{tab:intro} show, round-robin fails both \NEL\ and \AAU, the plain snake fails \NEL, and only the block--snake order satisfies all three. Note that \NEL\ is not the only conceivable notion of compensation, and \autoref{sec:nel} compares it with a symmetric alternative, the incomparability requirement.

Our results describe a trade-off between three goals: compensating those who receive fewer objects, limiting envy, and giving no one a reason to misreport. Our first set of results concerns the first two. We first characterize the \emph{draft rules}, in which agents take turns in a fixed order and each chooses her favorite remaining object: they are the only rules that satisfy \emph{non-wastefulness} (every object is allocated), \emph{neutrality} (relabeling the objects relabels the outcome) and \emph{object-wise Maskin monotonicity}, a weak form of Maskin's condition under which the allocation does not change when each agent's own objects only rise in her ranking (\autoref{thm:drafts}). The characterization fixes the agents and the objects and uses no incentive condition, and its proof reads the picking sequence off a profile at which all agents agree. Among draft rules, each fairness principle then fixes one feature of the order: \EFO\ the rounds, \NEL\ the precedence of the agents with fewer objects within each round, and \AAU\ the reversal of the order within each group. The block--snake drafts are therefore the only rules that satisfy the three principles together with the three structural conditions (\autoref{thm:strict}). We also show what a symmetric alternative to \NEL, the incomparability requirement, selects among draft rules, and we compare the orders under cardinal models (\autoref{sec:nel} and \autoref{app:add}).

Our second result concerns drafts with ties. A fan may be indifferent between some games, and in practice fans often want the top games of the season and are indifferent between the rest.\footnote{Indifferences in agents' preferences are also studied in papers on random matching \citep{BogomolnaiaMoulin2004}, two-sided matching \citep{ErdilErgin2017}, course allocation \citep{Li2020}, and housing markets \citep{Ehlers2026}.} Some games are ordered at the top, and we can treat the remaining games as tied below those listed at the top. Breaking such ties arbitrarily, for instance by drawing a random strict ranking, can cost efficiency. We propose a forward-looking version of the draft that instead resolves each tie in favor of the agents who pick later, and among them in favor of the earliest who strictly cares. For every picking sequence, its outcomes keep any guarantee of \EFO\ or \NEL\ that the draft provides, and they are efficient in the sense of pairwise dominance, or \emph{PD-efficient}, meaning that no other allocation gives every agent a bundle at least as good object by object and some agent one she strictly prefers (\autoref{thm:cf}). The block--snake draft thus remains fair and efficient when agents have ties.

Our last set of results shows that fairness conflicts with incentives. The rounds that \EFO\ requires separate each agent's turns by other agents' turns, and an agent can then gain by misreporting her ranking, passing over an object she expects to be still available at her next turn. Unless at most one agent receives more than one object, no draft rule that satisfies \EFO\ is even \emph{weakly strategy-proof}, whatever the number of agents (\autoref{prop:nelsp}). Weak strategy-proofness requires only that no misreport ever give an agent a bundle she strictly prefers object by object. With two agents who each receive at least two objects, the conflict extends beyond picking orders. Among non-wasteful neutral rules, we show that even weak strategy-proofness is incompatible with \EFO, already with four objects, whereas the known impossibility for strategy-proof rules requires five or more objects \citep{ABCM2017}. \EFO\ alone, without \NEL\ or \AAU, forces the conflict (\autoref{prop:imposs}). Alternation adds to the interleaving but is not needed for the cost. Compensation alone carries no such cost, although the only drafts that achieve it let each agent take all her objects at once, those with fewer objects first (\autoref{prop:nelsp}).

The closest characterizations are those of \citet{CB2026}, who characterize round-robin drafts by non-wastefulness, \EFO, resource monotonicity and respect of a given priority, and \citet{Aziz2026}, who characterizes fixed picking sequences in a model where the set of objects varies. We fix the agents and the objects, and in place of respect of a priority we use compensation and alternation, which select the block--snake rather than the round-robin sequence (\autoref{sec:lit}).

\autoref{sec:model} sets out the model, and \autoref{sec:fairness} discusses the compensatory fairness principles. \autoref{sec:draft} characterizes the block--snake draft and extends it to the cases with indifferences. \autoref{sec:incentives} studies incentives. \autoref{sec:lit} discusses related literature, and \autoref{sec:conclusion} concludes.

\section{Model}\label{sec:model}

A group of agents with equal claims divides a fixed set of indivisible objects, using only the rankings that the agents report.

\subsection{Agents and objects}

There is a finite set of \emph{agents} $N=\{1,\dots,n\}$, $n\ge 2$, and a finite set of \emph{objects} $O$ with
\[
|O| \;=\; T \;=\; kn+r
\]
for integers $k\ge0$ and $0\le r\le n-1$. Agents have equal claims, so no agent is entitled to more objects or to an earlier pick than another.

Each object is a distinct good (such as a ticket to a particular game), and any division of the objects into disjoint bundles is feasible.

\subsection{Preferences, dominance, efficiency}

Each agent ranks the objects, possibly with ties. Formally, let $\cW$ be the set of weak orders (complete and transitive binary relations) on $O$, and $\cP\subseteq\cW$ the set of linear orders, which have no ties. For $\succsim_i\in\cW$, we write $\succ_i$ and $\sim_i$ for its strict and indifference parts. A \emph{profile} is a list $\succsim=(\succsim_i)_{i\in N}\in\cW^N$ of the agents' orders. A profile in $\cP^N$ is \emph{strict}, and we then write $\succ$. Ties matter only from \autoref{sec:fl} on, and until then all profiles can be taken to be strict. A profile is \emph{unanimous} if all agents have the same order, and we use $\succsim^u$, or $\succ^u$ if it is strict, for both the profile and the common order. For a set $S\subseteq O$ and $\ell\le|S|$, $S(\ell)$ denotes the $\ell$-th most preferred element of $S$ under the order in question.

Agents compare bundles object by object, by \emph{pairwise dominance}, under which a set $S$ dominates a set $S'$ for agent $i$ if each object in $S'$ can be matched with a distinct object in $S$ that $i$ likes at least as much. Formally, for $S,S'\subseteq O$,
\[
S\ \dom{i}\ S' \quad\Longleftrightarrow\quad \text{there is an injection } \mu:S'\to S \text{ with } \mu(y)\succsim_i y \text{ for all } y\in S'.
\]
We write $S\sdom{i}S'$ if $S\dom{i}S'$ and not $S'\dom{i}S$, and $S\sim_iS'$ if both hold. In the literature this is called the responsive extension of $\succsim_i$ to bundles. It is incomplete, because some comparisons require knowing how much an agent prefers one object to another. Under $a\succ_ib\succ_ic\succ_id$, for instance, the bundles $\{a,d\}$ and $\{b,c\}$ are not comparable. For strict $\succ_i$, $S\dom{i}S'$ holds exactly when every additive utility with positive values that is consistent with $\succ_i$ ranks $S$ at least as high as $S'$. In our ordinal comparisons, saying that an agent does not envy a comparison bundle means that her own bundle pairwise-dominates it.

An \emph{allocation} is a list $x=(x_i)_{i\in N}$ of pairwise disjoint subsets of $O$, and $\cA$ is the set of allocations. We call $x_i$ agent $i$'s \emph{bundle}, and its size $|x_i|$ is the number of objects she receives. The allocation is \emph{non-wasteful} if every object is allocated, that is, $\bigcup_{i\in N}x_i=O$. An allocation $y$ \emph{PD-dominates} $x$ at $\succsim$ if $y_i\dom{i}x_i$ for all $i$ and $y_j\sdom{j}x_j$ for some $j$, that is, if every agent's bundle under $y$ is at least as good object by object and some agent strictly prefers hers. An allocation is \emph{PD-efficient} at $\succsim$ if no allocation PD-dominates it.\footnote{PD-efficiency is thus Pareto efficiency when agents compare bundles object by object. It does not rule out improvements that depend on how much agents prefer one object to another (\autoref{sec:commonpool}).} When discussing our ordinal allocation model, we use ``efficiency'' as shorthand for PD-efficiency unless another welfare criterion is stated.

\subsection{Rules}

A \emph{rule} assigns an allocation to every profile. It is a map $f:\cW^N\to\cA$ or, on the strict domain, $f:\cP^N\to\cA$.

Neutrality requires that the names of the objects play no role. To state it, rename the objects by a bijection $\sigma:O\to O$. For a profile $\succsim$, the profile $\succsim^\sigma$ ranks the renamed objects as $\succsim$ ranks the originals: $\sigma(a)\succsim^\sigma_i\sigma(b)\iff a\succsim_i b$. For an allocation $x$, $\sigma(x)=(\sigma(x_i))_{i\in N}$ gives each agent the renamed objects. A rule is \emph{neutral} if relabeling objects relabels the outcomes: $f(\succsim^\sigma)=\sigma(f(\succsim))$ for every bijection $\sigma$ of $O$.

Requirements on allocations, such as non-wastefulness, PD-efficiency, \EFO\ and \NEL, are stated at a profile $\succsim$. A rule satisfies a requirement if its outcome does at every profile.

\section{Compensatory fairness}\label{sec:fairness}

We impose envy-freeness up to one object:

\begin{description}[leftmargin=1.6em,style=nextline]
\item[\EFO\ (envy-freeness up to one object)] For all $i,j\in N$ there is $S\subseteq x_j$ with $|S|\le 1$ and $x_i\dom{i} x_j\setminus S$.
\end{description}
Any envy must disappear once a suitable single object is removed from the envied bundle. This is the ``envy bounded by a single good'' of \citet{Budish2011}, in the dominance version of \citet{CB2026}. Since a set cannot dominate a larger set, \EFO\ already bounds bundle sizes, and together with non-wastefulness it gives $r$ agents $k+1$ objects and the others $k$ (\autoref{lem:sizes} in \autoref{app:aux}). The unequal division is therefore a consequence of these two requirements, not an assumption of the model. Our two principles restrict how priority compensates for receiving fewer objects. An early pick may confer a lasting advantage only as compensation for receiving fewer objects. Between agents with different numbers of objects, priority must therefore favor those with fewer, round after round, since their disadvantage persists. This is \NEL\ (\autoref{sec:nel}). Among agents with the same number, no one is owed compensation, so an advantage at one step must be repaid at the next. This is \AAU\ (\autoref{sec:aau}).

\subsection{Compensation}\label{sec:nel}

\begin{description}[leftmargin=1.6em,style=nextline]
\item[\NEL\ (no envy toward larger bundles, up to any object)] For all $i,j\in N$ with $|x_i|<|x_j|$: $x_i\dom{i} x_j\setminus\{y\}$ for every $y\in x_j$.
\end{description}
Between an agent with fewer objects and one with more, \NEL\ asks more than \EFO, since removing \emph{any} single object from the larger bundle, not just a suitable one, must leave something the smaller agent does not envy. When the larger bundle has exactly one more object, this means that the smaller agent's bundle dominates the $|x_i|$ objects of the larger bundle that she prefers most.\footnote{Removing the least preferred object of $x_j$ leaves a set whose $\ell$-th most preferred element is at least as good, for $i$, as the $\ell$-th most preferred element of $x_j\setminus\{y\}$ for any $y\in x_j$; by \autoref{lem:sorted} it dominates every such set, so \NEL\ holds for all $y$ if and only if it holds for the least preferred one. If the larger bundle had two or more extra objects, \NEL\ would fail for size reasons alone (\autoref{lem:sizes}).} Between two agents who receive the same number of objects, at least two each, the same demand cannot always be met: at a unanimous profile, only one of them can hold the most preferred object in their combined bundles. For equal-sized bundles we therefore retain \EFO.

Why protect the agents with fewer objects more strongly than those with more, who receive only the protection of \EFO? The extra objects are a windfall, which the lottery hands to some agents and which no one is owed. \NEL\ requires the smaller bundle to dominate the larger one after any one object is removed. A comparison without removal cannot satisfy this demand, since a bundle can dominate only sets no larger than itself. With strict rankings, \NEL\ is equivalently a worst-case requirement. For every additive utility with positive values consistent with her ranking, an agent with fewer objects values her own bundle at least as much as a larger bundle with any one of its objects removed (\autoref{sec:model}). At a unanimous profile, the $k$ most preferred objects of the agent who holds the windfall are dominated by every bundle of an agent with fewer objects.

Because \NEL\ is ordinal, it cannot weigh the windfall against the advantage of picking first, and under some cardinal benchmarks those without a windfall come out ahead. With equally spaced values, for instance, Alice and Bob in the introduction's example score $10$ each and Carol $8$.\footnote{More generally, at a unanimous profile with equally spaced values $T,T-1,\dots,1$ the average agent with fewer objects finishes ahead of the average agent with more by $(kn-r-1)/2$ points under the block--snake draft.} We call a positive gap between the groups' mean normalized utilities \emph{overcompensation} of those with fewer objects; each agent's utility is normalized by her value for all objects divided by $n$. In the models and parameter cases of \autoref{app:cardinal}, the largest overcompensation occurs when all agents share the same steep values. Realized outcomes vary around these averages. In the two stochastic models, the block--snake draft has the lowest estimated mean absolute gap between the groups' mean utilities in every reported problem, although in the two-agent uniform model its margin over the plain snake and the Thue--Morse order is only about a tenth of a point. In the two reported problems with few objects per agent ($n=3$, $T=7$ and $n=4$, $T=19$), the group with the windfall is nevertheless ahead in 60 to 78 percent of draws even under the block--snake draft. These comparisons depend on the cardinal assumptions: heterogeneous preferences alone do not ensure a small gap or this ranking of the orders. As a worst-case ordinal requirement, \NEL\ gives a guarantee that holds whatever the windfall is worth. An alternative requirement, symmetric between the two groups and imposed at unanimous strict profiles, would ask that the smaller bundle and the larger bundle without its least preferred object be PD-incomparable: neither dominates the other under the common ranking. We call this the \emph{incomparability requirement}. It rules out exactly the domination that \NEL\ requires, so the two standards are incompatible there, and the incomparability requirement excludes the block--snake draft itself. It lets either side come out ahead, depending on the utilities. In the example of \autoref{tab:intro}, the plain snake satisfies the incomparability requirement but not \NEL, and it gives the windfall to Alice, who also picks first. Among draft rules, the incomparability requirement does not single out an order. When the objects do not divide evenly and the agents with fewer objects receive at most one object each, no non-wasteful rule satisfies it together with \EFO, because the sets it compares then have at most one element and are always comparable. When each agent receives at least two objects, together with \EFO\ and \AAU\ it admits a family of orders that includes the plain snake (\autoref{prop:neutralstd} in \autoref{app:add}). Among draft rules that satisfy \EFO\ and \AAU, the two standards thus select different orders when the division is uneven and each agent receives at least two objects: adding the incomparability requirement admits a family that includes the plain snake, whereas adding \NEL\ selects the block--snake draft.

When two fans share nineteen games, the block--snake draft lets the fan who receives nine games pick first in every round, so that under equally spaced values she scores $99$ and the fan with ten games $91$. The snake that starts with the nine-game fan gives each of them $95$. It satisfies \EFO\ and the incomparability requirement, but it fails \NEL, because the ten-game fan is ahead at the second step. The guarantee of \NEL\ can leave the recipient of the extra object behind in total utility when that object is worth little. Under the incomparability requirement, either group may have the higher value, depending on the utility profile. We accept the asymmetry of \NEL\ because the extra object is no one's due. One might object that the lottery already treats everyone alike, so no compensation after the draw is needed. But the members of a group see and compare the bundles they actually receive, not their chances, and \NEL\ is a guarantee about that realized division.

\subsection{Non-accumulation}\label{sec:aau}

Our second principle concerns priority, and we state it where priority is visible, at unanimous strict profiles. When all agents rank the objects the same way, differences between their bundles cannot come from differences in taste, so they reveal who had priority. At other profiles there is no common ranking by which to compare two agents' bundles step by step, and an agent can be ahead throughout because she likes objects that others do not want.

\begin{description}[leftmargin=1.6em,style=nextline]
\item[\AAU\ (alternating advantage under unanimity)] For a rule $f$ on $\cP^N$: at every unanimous profile $\succ^u$, the allocation $x=f(\succ^u)$ satisfies, for all $i,j\in N$ with $|x_i|=|x_j|$ and all $t<|x_i|$: if $x_i(t)\succ^u x_j(t)$, then $x_j(t+1)\succ^u x_i(t+1)$.
\end{description}
In words, when two agents with the same number of objects are compared by their most preferred objects, then by their second most preferred, and so on, whoever is ahead at one step is behind at the next. This reciprocity applies only to adjacent steps. The principle thus forbids priority from accumulating among equals, though it does not erase it, since the advantage at the last step is not repaid when agents with the same number of objects have an odd number of steps. It is vacuous when no two agents receive the same number of objects, as with two agents and an odd number of objects, and then only \NEL\ and \EFO\ constrain the outcome. The principle constrains only the order within a group of agents who receive the same number of objects. It does not say which group picks first, and with \NEL\ it does not imply \EFO\ when the division is uneven and each agent receives at least two objects (\autoref{cor:ef1}); the reversal it requires is also compatible with orders other than the block--snake (\autoref{prop:neutralstd} in \autoref{app:add}).

Local alternation is more than requiring advantages to balance out over time. Suppose that Alice and Bob share tickets to eight games, which both rank in the same order, numbered from $1$, the most preferred, to $8$, the least preferred, and call the agent with the higher-ranked object at a step the leader at that step. Three familiar picking orders give the allocations in \autoref{tab:aau}, where \emph{cumulative balance} requires that, at every step, the numbers of steps led by the two agents so far differ by at most one.
\begin{table}[ht]
\centering\small
\begin{tabular}{lcccc}
\toprule
Order & Alice, Bob & Leaders & \AAU & \shortstack{Cumulative\\balance}\\
\midrule
Round-robin ABABABAB & $\{1,3,5,7\}$, $\{2,4,6,8\}$ & AAAA & no & no\\
Balanced alternation ABBABAAB & $\{1,4,6,7\}$, $\{2,3,5,8\}$ & ABBA & no & yes\\
Block--snake ABBAABBA & $\{1,4,5,8\}$, $\{2,3,6,7\}$ & ABAB & yes & yes\\
\bottomrule
\end{tabular}
\caption{Two fans share eight tickets, numbered from $1$, the most preferred by both, to $8$, the least preferred. Round-robin repeats the same order every round, balanced alternation is the Thue--Morse order, and the block--snake reverses the order every round. Since both fans receive four tickets, the block--snake order is the plain snake here. At each step the leader is the fan with the higher-ranked ticket. \AAU\ requires the leader to change at every step, and cumulative balance requires only that the numbers of steps led never differ by more than one. Only the block--snake satisfies \AAU.}\label{tab:aau}
\end{table}
Round-robin lets Alice's first-move advantage persist at every step. The balanced alternation of \citet{BramsTaylor1999}, the Thue--Morse order, overcorrects.\footnote{In the computer-science literature on picking sequences, ``balanced alternation'' usually denotes the snake instead \citep{Celine2026}.} Bob leads at steps two and three, so Alice falls behind twice in a row before the balance is restored. \AAU\ accepts only the block--snake order.

The choice between the last two orders is normative. Balanced alternation equalizes more under some cardinal criteria. With eight tickets it equalizes the sums of squared ranks ($102$ each) as well as the rank sums, whereas the block--snake equalizes only the rank sums ($18$ each). \AAU\ uses no cardinal information, it can be checked at two adjacent steps, and it never lets an agent fall behind the same peer twice in a row. These features matter for groups that divide tickets by hand, where the snake is a common order: it needs no running tally of who has led so far, and each member can predict her next turn from her last.\footnote{Local alternation of this kind is also used in sports. Tennis tiebreaks serve in the repeated order ABBA, and football's rule-makers trialed the same order for penalty shootouts in 2017, prompted by evidence of a first-kicker advantage \citep{ApesteguiaPalaciosHuerta2010}, before dropping it in 2018; see \citet{Echenique2017} on the mechanism, \citet{KocherLenzSutter2012} on the disputed evidence, and \citet{BramsIsmail2018} for a catch-up rule, which lets whoever did worse in the previous round go first in the next.} Some contests depart from it. The National Fantasy Football Championship, a high-stakes fantasy contest, repeats the second round's order in the third, a ``third-round reversal'' that \AAU\ rules out because it lets the last picker of the first round lead in the second and third rounds. Such a correction is calibrated to a particular value curve,\footnote{The contest adopted the reversal when the first overall pick was a dominant advantage; see\newline\url{https://www.rotowire.com/football/article/football-draft-kit-28707}.} which an ordinal principle gives up in exchange for a guarantee that does not depend on how steeply values fall.

\section{The block--snake draft}\label{sec:draft}

This section introduces draft rules, characterizes them among all rules (\autoref{thm:drafts}), shows that the principles of \autoref{sec:fairness} single out the block--snake draft among them (\autoref{thm:strict}), and extends every draft to preferences with ties (\autoref{thm:cf}).

\subsection{Picking sequences and draft rules}

Drafts have been used to divide household property, elective courses and season tickets. They are easy to explain and to run, and each agent's outcome depends on the others' reports only through what is left at her turns.\footnote{A natural alternative is a pseudo-market. Approximate competitive equilibrium from equal incomes \citep{Budish2011} bounds envy by a single good at any market size and is approximately efficient and strategy-proof in large markets. It now allocates courses at Wharton, where it replaced a bidding auction \citep{BudishCachonKesslerOthman2017}. In the small groups we have in mind, its efficiency and incentive guarantees say little, and the market need not clear. The approximation error does not shrink with the number of agents, and since each ticket is a single unit, any error leaves a ticket unallocated or oversubscribed, in conflict with non-wastefulness. Its implementation, Course Match, also elicits cardinal reports and needs a computer to compute prices, whereas a draft needs only rankings and can be run by hand.}

A \emph{picking sequence} lists who picks at each turn. It is a vector $s=(s_1,\dots,s_T)\in N^T$, where $s_m$ is the agent who picks at turn $m$. On strict profiles the \emph{draft rule} $\Dr_s$ generated by $s$ lets the agents take turns in this order, each taking her favorite remaining object. Formally, given $\succ$, let $y_1$ be the element of $O$ that $s_1$ ranks first and, for $m=2,\dots,T$, let $y_m$ be the element of $O\setminus\{y_1,\dots,y_{m-1}\}$ that $s_m$ ranks first; then $\Dr_{s,i}(\succ)=\{y_m: s_m=i\}$. We call $y_m$ the \emph{$m$-th pick} and $s_m$ the \emph{$m$-th picker}.

The block--snake sequence formalizes the order of the introduction's example. Fix a bijection $\pi:N\to\{1,\dots,n\}$, the \emph{draw}, which ranks the agents and will be chosen by a uniform lottery (\autoref{sec:char}). The first $n-r$ agents in the draw form the block $A(\pi)=\{i:\pi(i)\le n-r\}$, which will receive $k$ objects, and the last $r$ form the block $B(\pi)=\{i:\pi(i)> n-r\}$, which will receive $k+1$. List $A(\pi)=\{a_1,\dots,a_{n-r}\}$ and $B(\pi)=\{b_1,\dots,b_r\}$ in increasing order of $\pi$. When $\pi$ is fixed we write $A$ and $B$. For $t\ge 1$ let $A^t=(a_1,\dots,a_{n-r})$ if $t$ is odd and $(a_{n-r},\dots,a_1)$ if $t$ is even, and define $B^t$ likewise, so that the order within each block reverses every round, as in a snake. The \emph{block--snake sequence} is
\[
s(\pi)\;=\;A^1\,B^1\;A^2\,B^2\;\cdots\;A^k\,B^k\;B^{k+1} ,
\]
of length $kn+r=T$. \emph{Round} $t\le k$ is the segment $A^tB^t$, and round $k+1$ is $B^{k+1}$. Each agent in $A(\pi)$ picks once in each of rounds $1,\dots,k$ and each agent in $B(\pi)$ once in each of rounds $1,\dots,k+1$, so an agent's $t$-th pick lies in round $t$. In the introduction's example, $A=\{\text{Alice},\text{Bob}\}$, $B=\{\text{Carol}\}$ and $k=2$, and $s(\pi)$ is Alice, Bob, Carol; Bob, Alice, Carol; Carol. The \emph{block--snake draft} on strict profiles is $\Dr^\pi=\Dr_{s(\pi)}$.

\subsection{Characterization}\label{sec:char}

We first characterize the draft rules among all rules, and then the block--snake draft among them. The first characterization uses, besides non-wastefulness and neutrality, a weak form of Maskin monotonicity \citep{Maskin1999}. \emph{Object-wise Maskin monotonicity} (\MM) says that if the objects each agent receives only move up in her ranking, the allocation does not change. Formally, a rule $f$ on $\cP^N$ satisfies \MM\ if $f(\succ')=x$ whenever $x=f(\succ)$ and $\succ'$ satisfies $\{z\in O: y\succ_i z\}\subseteq\{z\in O: y\succ'_i z\}$ for every $i\in N$ and every $y\in x_i$. We call such a $\succ'$ an \emph{\MM-improvement} of $x$. The condition is object-wise, and it is weaker than Maskin monotonicity with respect to preferences over bundles, which draft rules violate.\footnote{With the sequence $(i,j,i)$, objects $a,b,c$ and the rankings $a\succ_i b\succ_i c$ and $a\succ_j c\succ_j b$, agent~$i$ receives $\{a,b\}$. If she reports $b\succ a\succ c$ instead, $\{a,b\}$ is still her most preferred pair, so the bundles it dominates are unchanged, yet she receives $\{b,c\}$. \autoref{lem:decomp} in \autoref{app:add} decomposes the condition into two parts.} The condition lets the outcomes at unanimous profiles, where priority is visible, determine the outcome at every profile, and within this model it is what makes a rule a draft.

\begin{theorem}\label{thm:drafts}
A rule $f:\cP^N\to\cA$ is non-wasteful and neutral and satisfies \MM\ if and only if $f=\Dr_s$ for some picking sequence $s\in N^T$.
\end{theorem}

\begin{proof}
Every draft rule is non-wasteful and neutral and satisfies \MM\ (\autoref{lem:draft} and \autoref{lem:draftmm} in \autoref{app:aux}; the proofs of non-wastefulness and neutrality in \autoref{lem:draft} apply verbatim to any $\Dr_s$). Conversely, let $f$ satisfy the three conditions. Fix a unanimous profile with common order $z_1\succ^u\cdots\succ^u z_T$. By non-wastefulness every object is allocated; let $s_m$ be the agent who receives $z_m$. For any other common order $z'_1\succ^v\cdots\succ^v z'_T$, the bijection $\sigma(z_m)=z'_m$ satisfies $(\succ^u)^\sigma=\succ^v$, so by neutrality agent $s_m$ receives $z'_m$: the sequence $s$ does not depend on the unanimous profile. Now fix any $\succ\in\cP^N$, let $x=\Dr_s(\succ)$ with picks $y_1,\dots,y_T$, and let $\succ^u$ be the unanimous profile with common order $y_1\succ^u\cdots\succ^u y_T$. Then $f(\succ^u)$ gives $y_m$ to $s_m$, that is, $f(\succ^u)=x$. For $y_m\in x_{s_m}$, the objects below $y_m$ at $\succ^u$ are $y_{m+1},\dots,y_T$, and all of them lie below $y_m$ for $s_m$ at $\succ$, because they were available and not chosen at turn $m$. Hence $\succ$ is an \MM-improvement of $x$ relative to $\succ^u$, and $f(\succ)=x=\Dr_s(\succ)$.
\end{proof}

The theorem fixes the agents and the objects, uses no incentive condition, and yields every picking sequence, including interleaved ones.

Among draft rules, the fairness principles single out the block--snake draft, and with \autoref{thm:drafts} they characterize it among all rules.

\begin{theorem}\label{thm:strict}
\begin{enumerate}[label=(\alph*),nosep]
\item A draft rule $\Dr_s$ satisfies \EFO, \NEL\ and \AAU\ if and only if $s=s(\pi)$ for some draw $\pi$.
\item A rule $f:\cP^N\to\cA$ is non-wasteful and neutral and satisfies \EFO, \NEL, \AAU\ and \MM\ if and only if $f=\Dr^\pi$ for some draw $\pi$.
\end{enumerate}
\end{theorem}

The block--snake draft satisfies the three principles because each picker prefers her pick to anything picked later (\autoref{lem:draft} in \autoref{app:aux}). Since every agent's $t$-th pick comes before anyone's $(t+1)$-th, this gives \EFO. Since in each round the agents with fewer objects pick before the others, it also gives \NEL, and the reversal of the order within each block from round to round gives \AAU.

For necessity, it suffices to consider unanimous profiles, where a draft rule gives the $m$-th ranked object to the $m$-th picker, so each principle becomes a condition on the picking sequence:
\[
\begin{array}{lcl}
\EFO & \Longrightarrow & \text{the picks form rounds, everyone's $t$-th pick before anyone's $(t+1)$-th;}\\[2pt]
\NEL & \Longrightarrow & \text{in every round, the agents with $k$ objects pick before those with $k+1$;}\\[2pt]
\AAU & \Longrightarrow & \text{the order within each block reverses from one round to the next.}
\end{array}
\]
Each principle restricts one feature of the order, and in general none is implied by the other two (\autoref{cor:ef1} below and \autoref{prop:indep} in \autoref{app:add}).

Within the draft rules, \AAU\ alone already implies the round structure among agents with the same number of objects, so \EFO\ is needed only between agents with different numbers of objects, and then only when some agent receives at least three:

\begin{corollary}\label{cor:ef1}
For given $(n,k,r)$, every draft rule that satisfies \NEL\ and \AAU\ satisfies \EFO\ if and only if $r=0$ or $k\le1$.
\end{corollary}

\autoref{thm:strict}(b) follows from \autoref{thm:strict}(a) and \autoref{thm:drafts}. The proof in \autoref{app:proofs} shows that the three principles need to hold only at unanimous profiles, where they force the sequence that \autoref{thm:drafts} reads off to be $s(\pi)$.

Our principles thus determine the rule only up to the draw $\pi$, which decides who receives $k+1$ objects and the order of picks. Because agents have equal claims, the \emph{random block--snake draft} chooses $\pi$ uniformly at random and then applies $\Dr^\pi$, so the symmetry among agents before the draw comes from the lottery, not from the principles. Renaming the agents only permutes the draw, so every agent has the same chance of occupying each position in it, and hence of receiving $k+1$ objects. This equal treatment concerns chances, not expected welfare, since the agents' preferences differ.

\subsection{Indifferences: the forward-looking draft}\label{sec:fl}

Agents are often indifferent between objects. A fan may not care which of several weekday games she attends. With indifferences, ``take your favorite remaining object'' is ambiguous, and how the ambiguity is resolved matters for others. An agent indifferent between $a$ and $b$ who takes $a$ may block a later agent who strictly wants $a$. Breaking ties in advance, by fixed or random strict rankings, can cost efficiency (\autoref{ex:two}).

With ties and at least two objects, no single-valued rule is non-wasteful, envy-free up to one object and neutral (\autoref{prop:noneutral} in \autoref{app:add}), because when every agent is indifferent among all objects, a neutral rule has no way to break ties. We therefore work with \emph{correspondences}, maps $\Delta:\cW^N\to2^{\cA}\setminus\{\emptyset\}$ that may select several allocations. A \emph{selection} from $\Delta$ is a rule $g$ with $g(\succsim)\in\Delta(\succsim)$ for every $\succsim$. A correspondence satisfies a requirement on allocations if every allocation it selects does, and it is \emph{neutral} if $\Delta(\succsim^\sigma)=\{\sigma(x):x\in\Delta(\succsim)\}$ for every bijection $\sigma$ of $O$.\footnote{A correspondence that selects a single allocation at every strict profile satisfies \AAU\ if its restriction to $\cP^N$, which is then a rule, does. With ties, the $\ell$-th most preferred element $S(\ell)$ of a set $S$ is determined only up to indifference, but its indifference class is well defined, and every comparison involving $S(\ell)$ depends only on that class.}

The forward-looking draft $L_s$ generated by a picking sequence $s$ is defined turn by turn. At each turn the picker is committed to her favorite objects among those that the commitments of earlier turns can still spare, and $L_s(\succsim)$ consists of the allocations that give every turn an object to which its picker is committed, using no object twice. \autoref{app:aux} states the definition in full. We write $L^\pi=L_{s(\pi)}$. The construction extends to picking sequences the priority-based rules that \citet{Svensson1994} and \citet{BDE2005} study on the full preference domains of their assignment models, and the proof of PD-efficiency adapts the standard argument for serial dictatorship. Whether an object can still be spared is a bipartite matching question, so an allocation of the forward-looking draft can be computed efficiently (\autoref{app:add}).

\begin{example}\label{ex:two}
Two agents pick in the order $(1,2)$ from $O=\{\text{Liverpool},\text{City}\}$, tickets to games against the two clubs. If both agents are indifferent, each is committed to both tickets, and $L_s$ contains both allocations; the agents are indifferent between them, so the forward-looking draft is single-valued only up to indifference. If agent $1$ is indifferent and agent $2$ strictly prefers Liverpool, agent $1$ is committed to both tickets, so Liverpool can still be spared for agent $2$, who is committed to it, and agent $1$ receives City. Her indifference is resolved in favor of the agent who cares. Breaking agent $1$'s tie in favor of Liverpool instead would give agent $2$ City, an allocation that exchanging the two tickets improves for agent $2$ without harming agent $1$.
\end{example}

The forward-looking draft extends every draft rule to all weak orders on objects. Unlike a live draft, in which agents choose in real time, leave ties to their own discretion and may pick strategically (\autoref{sec:conclusion}), the forward-looking draft is a direct mechanism that needs all rankings before the first pick. It coincides with the draft rule when no agent has ties, and at every profile it keeps the draft's fairness guarantees and its PD-efficiency:

\begin{theorem}\label{thm:cf}
Fix a picking sequence $s$ and a profile $\succsim\in\cW^N$.
\begin{enumerate}[label=(\alph*),nosep]
\item $L_s(\succsim)$ is nonempty, and every allocation in it is non-wasteful and PD-efficient.
\item If $\Dr_s$ satisfies \EFO, every allocation in $L_s(\succsim)$ satisfies \EFO\ at $\succsim$, and if $\Dr_s$ satisfies \NEL, every allocation in $L_s(\succsim)$ satisfies \NEL\ at $\succsim$.
\item Any two allocations in $L_s(\succsim)$ give every agent bundles she is indifferent between; $L_s$ is neutral; and $L_s(\succ)=\{\Dr_s(\succ)\}$ at every strict profile.
\end{enumerate}
\end{theorem}

In particular, at every profile, every allocation of the forward-looking block--snake draft $L^\pi$ is non-wasteful, PD-efficient and envy-free up to one object and satisfies \NEL, with $n-r$ agents receiving $k$ objects and $r$ agents receiving $k+1$. When no agent has ties, $L^\pi$ coincides with $\Dr^\pi$, which \autoref{thm:strict} characterizes.

With ties, the random block--snake draft applies $L^\pi$ in place of $\Dr^\pi$, and every outcome of the lottery keeps the properties of \autoref{thm:cf}. Which allocation in $L^\pi(\succsim)$ is chosen does not matter to any agent, since by \autoref{thm:cf}(c) every agent is indifferent between them. If equal treatment is also wanted for the physical objects agents receive, the allocation can itself be drawn uniformly from $L^\pi(\succsim)$.

On strict profiles every draft rule is PD-efficient, and the content of \autoref{thm:cf} is that the forward-looking resolution of ties preserves this, together with any guarantee of \EFO\ or \NEL. When several tie-breaks are efficient, the forward-looking draft favors the earliest later picker who cares:

\begin{example}\label{ex:three}
Three agents pick in the order $(1,2,3)$ from $\{a,b,c\}$, with $\succsim_1:a\sim b\succ c$, $\succ_2:a\succ c\succ b$ and $\succ_3:b\succ a\succ c$. Both ways of breaking agent $1$'s tie give PD-efficient allocations, $(b,a,c)$ and $(a,c,b)$, where the $i$-th entry is agent $i$'s object. The forward-looking draft selects $(b,a,c)$: agent $1$'s indifference is resolved in favor of agent $2$, who picks before agent $3$.
\end{example}

Any way of breaking ties already preserves the block--snake draft's fairness guarantees: running $s(\pi)$ at a strict refinement of the reported profile, obtained by breaking ties without changing any strict comparison, yields an allocation that satisfies \EFO\ and \NEL\ with respect to the reported weak orders (\autoref{lem:nest} in \autoref{app:aux}). What the forward-looking draft adds is PD-efficiency, which arbitrary tie-breaking can lose (\autoref{ex:two}), a definite priority in resolving ties, and the properties in \autoref{thm:cf}(c). The priority has a cost for an agent who is only partly indifferent, because it favors the next turn, which may belong to another agent, over her own later turns.

\begin{example}\label{ex:cost}
Two agents pick in the order $(1,2,2,1)$ from $O=\{a,b,c,d\}$, with $\succsim_1:a\sim b\succ c\succ d$ and $\succ_2:a\succ c\succ d\succ b$. The forward-looking draft resolves agent $1$'s tie in favor of agent $2$'s next turn: agent $2$ takes $a$ and then $c$, and agent $1$ receives $\{b,d\}$. The allocation $(\{a,b\},\{c,d\})$, which agent $1$ strictly prefers, is also PD-efficient, and she would obtain it by reporting $a\succ b$.
\end{example}

Whether a resolution that favors the tie-holder's own later turns preserves the properties of \autoref{thm:cf} is open. A fan whose only ties are in a bottom indifference class, as when she reports a top list and leaves the remaining games unranked, is not affected by the priority. She keeps the same ranked games under every strict refinement of that class, because until her first pick from that class the commitments are unchanged, by then none of her ranked games remains feasible, and later feasible sets can only shrink; her ties among the unranked games are resolved in favor of the later pickers who strictly care, and if the indifferences are genuine she is indifferent between the resulting bundles. For fixed strict reports by the other agents, every outcome from a tied report can be reproduced by some strict refinement of that report (\autoref{lem:nest} in \autoref{app:aux}). The refinement may depend on the other agents' reports; the observation does not identify one refinement that is optimal against every possible profile, nor does it make the rule strategy-proof.

\section{Incentives}\label{sec:incentives}\label{sec:commonpool}

A fan should be able to say which games she wants most without worrying that honesty will cost her. Call a rule $f$ on $\cP^N$ \emph{strategy-proof} if $f_i(\succ)\dom{i}f_i(\succ'_i,\succ_{-i})$ for all $\succ$, $i$ and $\succ'_i$, that is, if reporting truthfully gives every agent a bundle at least as good, object by object, as any misreport. It is \emph{weakly strategy-proof} if there are no $\succ$, $i$ and $\succ'_i$ with $f_i(\succ'_i,\succ_{-i})\sdom{i}f_i(\succ)$, that is, if no misreport gives an agent a bundle she strictly prefers object by object. Because pairwise dominance is incomplete, the second requirement is weaker, since it allows misreports that lead to bundles the agent cannot compare with her truthful one.

Draft rules in which an agent's picks are separated by other agents' picks are manipulable \citep{BouveretLang2014}. An agent who knows that her favorite object will still be available at her next turn can take a contested object first and collect her favorite later (\autoref{lem:manip} in \autoref{app:add}). If at most one agent receives more than one object, that is, if $k=0$, or $k=1$ and $r\le1$, every agent's picks in $s(\pi)$ are consecutive and $\Dr^\pi$ is strategy-proof. In every other configuration $\Dr^\pi$ is not even weakly strategy-proof. Since $L^\pi$ coincides with $\Dr^\pi$ on strict profiles, the same manipulation applies to $L^\pi$ and to every selection from it.\footnote{Weaker incentive notions for set-valued rules do not help, since the manipulation uses strict profiles, at which $L^\pi$ selects a single allocation. The text also fixes the draw, but manipulability survives when reports must be submitted before the draw is made. Let $n=2$ and $O=\{a,b,c,d,e\}$, so that the two draws give the sequences $(1,2,1,2,2)$ and $(2,1,2,1,1)$, and let $\succ_1:a\succ b\succ c\succ d\succ e$ and $\succ_2:a\succ c\succ d\succ b\succ e$. Truthfully, agent~$1$ receives $\{a,b\}$ under the first draw and $\{b,d,e\}$ under the second. If she reports $\succ'_1:a\succ c\succ b\succ d\succ e$, she still receives $\{a,b\}$ under the first draw. Under the second, she takes $c$ at her first turn and still finds $b$ at her next, because agent~$2$ prefers $d$ to $b$, so she receives $\{b,c,e\}$, which strictly dominates $\{b,d,e\}$. The misreport never hurts her and sometimes helps, so it pays whatever the chances of the two draws.}

The compensation principle by itself creates no incentive problem, but the rounds that \EFO\ requires do. A \emph{quota serial dictatorship} is the draft rule of a sequence in which every agent's picks are consecutive, so that agents choose their entire bundles one after another, each taking her favorite remaining objects up to her \emph{quota}, the number of objects she is to receive. Since no one else picks between an agent's choices, the rule is strategy-proof, and, being a draft rule, it is non-wasteful, neutral and PD-efficient and satisfies \MM.

\begin{proposition}\label{prop:nelsp}
\begin{enumerate}[label=(\alph*),nosep]
\item A draft rule satisfies \NEL\ and is weakly strategy-proof if and only if it is a quota serial dictatorship in which $n-r$ agents have quota $k$, $r$ agents have quota $k+1$, and every agent with quota $k$ chooses before every agent with quota $k+1$, the orders within the two groups being arbitrary. In particular, for every $(n,k,r)$ there is a non-wasteful, neutral, PD-efficient and strategy-proof rule that satisfies \MM\ and \NEL.
\item If at least two agents receive more than one object, no draft rule that satisfies \EFO\ is weakly strategy-proof.
\end{enumerate}
\end{proposition}

Among picking orders, strategy-proofness requires each agent's picks to be consecutive. Quota serial dictatorships are, under standard conditions, the strategy-proof rules of multi-unit assignment \citep{Papai2001,Hatfield2009}, and part (a) identifies which of them are compensatory. \NEL\ alone is thus a weak requirement, since in such a rule the first agent with a small quota takes her $k$ favorite objects at once. Part (b) holds for any number of agents. \EFO\ requires rounds, and whenever at least two agents receive more than one object, some agent's picks in consecutive rounds are separated by another agent's pick, which is what \autoref{lem:manip} exploits. When all agents receive equally many objects, part (b) also follows from \citet[Theorem~1]{Caspari2026}, since every draft rule is PD-efficient, and with two objects each from \citet[Theorem~3]{CB2026}. Its new content is the case in which the objects do not divide evenly.

Interleaving also affects efficiency. Every draft rule is PD-efficient \citep{BramsKing2005,CB2026}, as the proof of \autoref{thm:cf} also shows, but PD-efficiency is weak, since it rules out only improvements that every agent can see object by object. Among allocations with the same bundle sizes, quota serial dictatorships are Pareto efficient for every profile of additive utilities consistent with the rankings \citep{Hatfield2009,Caspari2026}, whereas the block--snake draft is not. When two agents pick in the order $1,2,2,1$ under the common order $a\succ b\succ c\succ d$, they receive $\{a,d\}$ and $\{b,c\}$, and under the utilities $(4,3,2,0.5)$ for agent~$1$ and $(100,3,2,1)$ for agent~$2$, listed for $a,b,c,d$, exchanging the bundles benefits both. Quota serial dictatorships thus combine this efficiency with strategy-proofness but violate \EFO\ whenever at least two agents receive more than one object.

The conflict also arises outside the class of draft rules. For two agents with additive valuations, \citet[Application~4.6]{ABCM2017} show that no strategy-proof rule that allocates every object satisfies \EFO\ once there are five or more objects.\footnote{Break ties in values by a fixed order of the objects. An ordinal rule that is strategy-proof with respect to pairwise dominance then induces a mechanism that is truthful for every additive valuation, and an allocation that satisfies \EFO\ with respect to pairwise dominance satisfies it for every additive valuation consistent with the rankings, so their result applies here.} Among neutral rules, weak strategy-proofness already suffices, and the conflict starts at four objects. It is forced by \EFO\ alone, without \NEL\ or \AAU:

\begin{proposition}\label{prop:imposs}
Let $n=2$ and $k\ge2$. No non-wasteful, neutral and weakly strategy-proof rule $f:\cP^N\to\cA$ satisfies \EFO.
\end{proposition}

The proof uses the fact that, with two agents, a profile at which the agents report different orders is a single deviation away from two unanimous profiles, where neutrality fixes the outcome. Neutrality is needed only at these unanimous profiles, but it cannot be dropped at four objects. In the rule in which agent~$1$ takes her preferred object from a fixed pair, agent~$2$ takes her preferred object from the other pair, and each remainder goes to the other agent, both agents receive two objects, and the rule is strategy-proof and satisfies \EFO\ \citep{ABCM2017}, but it is not neutral. For two agents, the proposition confirms among neutral rules a conjecture of \citet{CB2026} that non-wastefulness, \EFO\ and weak strategy-proofness are incompatible. With three or more agents the argument fails, because the reports of the remaining agents can affect how the others' objects are divided. Related impossibilities hold for any number of agents, for nonbossy rules (an agent cannot change others' bundles without changing her own; \autoref{app:incentives}) and, when all agents receive equally many objects, for PD-efficient rules \citetext{\citealp[Theorem~3]{CB2026}; \citealp[Theorem~1]{Caspari2026}}. Beyond picking orders, the condition of \autoref{prop:nelsp}(b) does not suffice. With three or more agents, some of whom receive only one object, a non-wasteful, neutral and strategy-proof rule can satisfy \EFO\ (\autoref{rem:kone} in \autoref{app:incentives}). Whether \autoref{prop:imposs} extends to three or more agents when every agent receives at least two objects is open (\autoref{sec:conclusion}).

For every positive additive utility consistent with an agent's ranking, truth-telling in any draft rule maximizes both her worst-case and her best-case utility over the others' reports (\autoref{prop:nom} in \autoref{app:incentives}). A rule that makes truth-telling maximize worst-case utility is called \emph{maxmin strategy-proof}. Since truth-telling also maximizes best-case utility, every draft rule is \emph{not obviously manipulable} in the sense of \citet{TroyanMorrill2020}. An agent who judges reports by either criterion therefore has no reason to misreport. A misreport can still pay off against many different reports of the others, so an agent need not know them exactly to gain. In the Harvard Business School course draft, students nevertheless manipulated substantially and at a material welfare cost \citep{BudishCantillon2012}.

The manipulations above exploit the fact that, in a common pool, each pick changes what is available to everyone at later turns. When the objects instead come in separate types, such as games grouped by month, and fairness is judged type by type, the same priority pattern can be implemented by a strategy-proof mechanism (\autoref{sec:separable}).

\section{Related literature}
\label{sec:lit}
Drafts are the standard sequential mechanism for multi-unit assignment \citep{BudishCantillon2012,BouveretLang2014,BKP2022,Caspari2026}. \citet{CB2026} characterize round-robin draft rules, in which agents pick one object per round in the order of a given priority, using resource monotonicity, which requires that adding objects never make an agent worse off. An earlier version characterizes them without taking a priority as given, in a model with variable populations \citep{CB2024}. We instead fix the population and the objects, let agents receive different numbers of objects, and use object-wise Maskin monotonicity to recover a picking sequence from the allocations. Our compensation principle, \NEL, is a priority-free analogue of their respect of a priority, under which no agent envies an agent of lower priority. Under \NEL, the agents with fewer objects take the place of the higher-priority agents, and envy toward a larger bundle is ruled out up to any one object. \autoref{thm:drafts} adapts the classical argument for serial dictatorship \citep{Svensson1999}, which reads the order off a unanimous profile and extends it by monotonicity. The multi-unit characterizations of \citet{Papai2001}, \citet{EhlersKlaus2003} and \citet{Hatfield2009} rest on strategy-proofness and yield serial or sequential dictatorships, whereas \autoref{thm:drafts} uses only object-wise Maskin monotonicity, non-wastefulness and neutrality and yields every picking sequence, including interleaved ones. Maskin monotonicity \citep{Maskin1999} characterizes several rules that assign one object to each agent \citep{Takamiya2001,KojimaManea2010}. With one object per agent, the object-wise and bundle-wise versions coincide.

The round-based, or recursively balanced, sequences, in which every agent picks once per round, are exactly the picking sequences that guarantee \EFO, and among them, the sequences with the best maximin-share guarantee include those that let the agent who picks last in the first round pick first in every later round \citep{Celine2026}. Those sequences compensate for position in the first round, whereas ours compensate for the number of objects. We ask the reverse question of which principles on allocations select a sequence. In independent work, \citet{Aziz2026} characterizes the rules generated by a fixed picking sequence by bundle-size independence and resource monotonicity, in a model where the set of objects varies, and shows that adding \EFO\ yields the round-based sequences. On a fixed set of objects, \autoref{thm:drafts} reaches the fixed sequences through non-wastefulness, neutrality and object-wise Maskin monotonicity, and \autoref{thm:strict} selects the block--snake sequence among them. In quota serial dictatorships, giving later choosers more objects is treated as a design choice and resolved by optimizing welfare \citep{BouveretGilbert2025}. The efficiency cost of breaking ties exogenously is familiar from school choice, where lotteries break ties in schools' priorities \citep{ErdilErgin2008}.

Fairness with unequal bundle sizes has been studied mainly when agents have unequal entitlements \citep{CISZ2021,CSS2021}, and the question of who receives a remainder is the classical question of apportionment \citep{BalinskiYoung1982}.

The conflict between envy-freeness up to one object and incentives is known in related settings; \autoref{sec:incentives} and \autoref{app:incentives} relate \autoref{prop:nelsp} and \autoref{prop:imposs} to the impossibilities of \citet{ABCM2017}, \citet{CB2026} and \citet{Caspari2026}. \citet{Caspari2026} also proposes booster drafts, which run a serial dictatorship in each of several separate pools and reverse the priority order from one pool to the next; \autoref{sec:separable} gives their reversal a normative reading and extends them to a short pool.

\section{Conclusion}\label{sec:conclusion}

When people with equal claims divide objects that cannot be split, some must receive more than others, and without money the only available compensation for those who receive fewer may be to let them pick first. In this paper, we have asked how far such priority should go to ensure fairness. Two compensatory fairness principles help to answer this question: no agent with fewer objects should envy a larger bundle with any one object removed (\NEL), and among agents with equally many objects, a head start should be repaid at the next step (\AAU). Together with envy-freeness up to one object (\EFO), they single out the block--snake draft among picking orders, and its forward-looking version keeps these guarantees, together with PD-efficiency, when agents have ties, including top-list reports completed by treating unranked objects as those tied at the bottom. 

For a group dividing heirlooms or season tickets, the recommendation is simple. Draw lots to decide who receives the extra objects and the order within each group, let those without an extra object pick first in every round, and reverse the order within each group from one round to the next. There is an incentive cost, but the cost comes from limiting envy, not from compensation. Among draft rules, compensation alone is compatible with strategy-proofness, but only through a quota serial dictatorship, in which those with fewer objects choose all their objects first (\autoref{prop:nelsp}). Limiting envy requires rounds, and in rounds an agent can gain by passing over an object she expects to remain available. This holds for every picking order that limits envy in this way whenever two or more agents receive more than one object (\autoref{prop:nelsp}) and, when there are two agents who each receive at least two objects, for every non-wasteful and neutral rule (\autoref{prop:imposs}).

Three questions remain open and provide an avenue for future research. The first question is the mirror problem of rotas, such as on-call schedules or allocation of teaching loads in a department, where receiving fewer (duties) is the advantage. Compensation then runs the other way, and since the duties left at the end of a draft are the least preferred ones, when the extra duties should be chosen is itself unclear. The second question is whether the incentive cost extends beyond picking orders when there are three or more agents and every agent receives at least two objects. When some agents receive only one object it does not (\autoref{rem:kone}), and the two-agent argument does not extend, because a third agent's report can affect how the other agents' objects are divided. The third question is whether compensation survives strategic play. With two agents who choose objects in turn under complete information, the picking game, whose equilibria \citet{KohlerChandrasekaran1971} first constructed, has a unique subgame-perfect equilibrium allocation \citep{Kalinowski2013}, and for round-robin, pure Nash equilibria of the reporting game are envy-free up to one object with respect to true values \citep{Amanatidis2023}. Whether the corresponding allocations satisfy \EFO\ and \NEL\ is open.

\appendix

\section{Auxiliary results}\label{app:aux}

In the appendices we abbreviate non-wastefulness by \NW, neutrality by \NEU, PD-efficiency by \EFF, strategy-proofness by \SP, weak strategy-proofness by \WSP\ and nonbossiness by \NB. A correspondence is \emph{single-valued on strict profiles} (\textsf{SV}) if it selects exactly one allocation at every strict profile. Following \citet{BDE2005}, a correspondence is \emph{essentially single-valued} (\textsf{ESV}) if every agent is indifferent between any two allocations it selects, that is, $x,y\in\Delta(\succsim)$ implies $x_i\sim_iy_i$ for every $i$; it is \emph{Pareto indifferent} (\textsf{PI}) if it also selects every allocation that all agents find equally good, that is, $x\in\Delta(\succsim)$, $y\in\cA$ and $y_i\sim_ix_i$ for every $i$ imply $y\in\Delta(\succsim)$.

\begin{lemma}\label{lem:sorted}
Let $\succsim$ be a weak order on $O$ and $S,S'\subseteq O$ with $|S|\ge|S'|$. Then $S\succsim^{\PD}S'$ if and only if $S(\ell)\succsim S'(\ell)$ for every $\ell\le|S'|$: a set dominates a set that is no larger exactly when its $\ell$-th most preferred element is at least as good as the other's, for every $\ell$.
\end{lemma}

\begin{proof}
If $S(\ell)\succsim S'(\ell)$ for every $\ell$, then $S'(\ell)\mapsto S(\ell)$ is the required injection. Conversely, suppose $S'(\ell)\succ S(\ell)$ for some $\ell$. Under any injection $\mu$ with $\mu(y)\succsim y$, the $\ell$ elements $S'(1),\dots,S'(\ell)$, all of which are at least as good as $S'(\ell)$ and hence ranked strictly above $S(\ell)$, are mapped to $\ell$ distinct elements of $S$ ranked strictly above $S(\ell)$. But $S$ has at most $\ell-1$ such elements, a contradiction.
\end{proof}

\begin{lemma}\label{lem:sizes}
Let $\succsim\in\cW^N$ and let $x$ satisfy \EFO\ or \NEL\ at $\succsim$. Then $\big||x_i|-|x_j|\big|\le1$ for all $i,j\in N$. If $x$ is also non-wasteful, exactly $r$ agents receive $k+1$ objects and the others $k$.
\end{lemma}

\begin{proof}
If $|x_j|\ge|x_i|+2$, then for every $S\subseteq x_j$ with $|S|\le1$ the set $x_j\setminus S$ has more elements than $x_i$, so no injection from it into $x_i$ exists, and both \EFO\ and \NEL\ fail for the pair $(i,j)$. Now let $x$ also be non-wasteful, and let $m$ be the smallest bundle size. The sizes lie in $\{m,m+1\}$, fewer than $n$ agents receive $m+1$ objects, and the sizes sum to $T=kn+r$ with $0\le r\le n-1$, so uniqueness of division with remainder gives $m=k$ and exactly $r$ agents of size $k+1$.
\end{proof}

\begin{lemma}\label{lem:draft}
Fix a picking sequence $s$ and $\succ\in\cP^N$, and let $y_1,\dots,y_T$ be the picks of $\Dr_s$ with pickers $s_1,\dots,s_T$.
\begin{enumerate}[label=(\alph*)]
\item $y_m\succ_{s_m} y_{m'}$ whenever $m<m'$.
\item If $s=s(\pi)$, then $\Dr^\pi=\Dr_s$ satisfies \NW, \EFO, \NEL, \AAU\ and \NEU.
\end{enumerate}
\end{lemma}

\begin{proof}
(a) At turn $m$ the object $y_{m'}$ ($m'>m$) is still available and is not chosen, so $y_m\succ_{s_m}y_{m'}$.

(b) Let $x=\Dr_s(\succ)$. \NW\ is immediate. Write $i_1,i_2,\dots$ for the picks of agent $i$ in order of time, so $i_t$ is $i$'s $t$-th pick and lies in round $t$.

\emph{\EFO.} Let $i,j\in N$. If $x_j=\emptyset$, take $S=\emptyset$. Otherwise, since $i$'s $t$-th pick is made in round $t$ and $j$'s $(t+1)$-th pick in round $t+1$, (a) gives $i_t\succ_i j_{t+1}$ for all $t\le|x_j|-1$; as $|x_i|\ge|x_j|-1$, the map $j_{t+1}\mapsto i_t$ is an injection from $x_j\setminus\{j_1\}$ into $x_i$ whose images $i$ weakly prefers.

\emph{\NEL.} Let $i\in A(\pi)$ and $j\in B(\pi)$ (these are the only pairs with $|x_i|<|x_j|$). In each round $t\le k$ the $A$-block precedes the $B$-block, so $i$'s $t$-th pick precedes every pick of $j$ from round $t$ on; by (a), $i_t\succ_i j_{t'}$ for all $t'\ge t$. For any $\ell\in\{1,\dots,k+1\}$ define $\mu$ on $x_j\setminus\{j_\ell\}$ by $\mu(j_{t'})=i_{t'}$ for $t'<\ell$ and $\mu(j_{t'})=i_{t'-1}$ for $t'>\ell$. The images $i_1,\dots,i_{\ell-1},i_\ell,\dots,i_k$ are distinct, and in both cases the image is $i_t$ for some $t\le t'$, so $\mu(j_{t'})\succ_i j_{t'}$ by the observation above. Hence $x_i\dom{i}x_j\setminus\{j_\ell\}$.

\emph{\AAU.} At a unanimous profile the highest-ranked remaining object is always the next object of the common order, so $\Dr^\pi$ picks the objects in the order $\succ^u$ and $x_i(t)=i_t$. Agents of equal size belong to the same block. If $i_t\succ^u j_t$ then $i$ precedes $j$ in the block order of round $t$; the block order of round $t+1$ is the reverse (this includes rounds $k$ and $k+1$ for block $B$), so $j$ precedes $i$ there and $j_{t+1}\succ^u i_{t+1}$.

\emph{\NEU.} By induction on $m$, the picks of $\Dr^\pi$ at $\succ^\sigma$ are $\sigma(y_1),\dots,\sigma(y_T)$ with the same pickers.
\end{proof}

\begin{lemma}\label{lem:draftmm}
For every picking sequence $s$, the draft rule $\Dr_s$ satisfies \MM.
\end{lemma}

\begin{proof}
Let $\succ'$ be as in the definition of \MM\ with $x=\Dr_s(\succ)$, and let $y_1,\dots,y_T$ be the picks of $\Dr_s$ at $\succ$. By induction on $m$, the first $m-1$ picks at $\succ'$ coincide with those at $\succ$, so the available set at turn $m$ is the same; for every available $z\neq y_m$ we have $y_m\succ_{s_m} z$ by the choice of $y_m$, hence $y_m\succ'_{s_m}z$ by the hypothesis applied to $y_m\in x_{s_m}$. So $y_m$ is again chosen.
\end{proof}

The forward-looking draft of \autoref{sec:fl} is defined as follows. Fix a picking sequence $s$ and a profile $\succsim\in\cW^N$. The \emph{commitment sets} $C_1,\dots,C_T\subseteq O$ are defined recursively, turn by turn. For $m\in\{1,\dots,T+1\}$, an \emph{$m$-partial assignment} fills the first $m-1$ turns with distinct objects, each from its commitment set: it is an injection $\varphi:\{1,\dots,m-1\}\to O$ with $\varphi(j)\in C_j$ for every $j<m$ (for $m=1$ the empty map). For $m\le T$, the \emph{feasible set} of turn $m$ is
\[
F_m\;=\;\{z\in O:\ \text{some $m$-partial assignment avoids }z\},
\]
where avoiding $z$ means not using it. It contains the objects that the earlier commitments can still spare, and the commitment set $C_m$ is the set of $\succsim_{s_m}$-maximal elements of $F_m$, the picker's favorites among them. Write $K_m$ for the $\succsim_{s_m}$-indifference class containing $C_m$, so that $C_m=K_m\cap F_m$. We compare two indifference classes of the same agent by any representatives. The \emph{forward-looking draft} generated by $s$ is the correspondence
\[
L_s(\succsim)\;=\;\big\{x^\varphi:\ \varphi\text{ is an injection from }\{1,\dots,T\}\text{ to }O\text{ with }\varphi(m)\in C_m\text{ for all }m\big\},
\]
where $x^\varphi_i=\varphi(\{m:s_m=i\})$ gives each agent the objects assigned to her turns, and $L^\pi=L_{s(\pi)}$.

\begin{lemma}\label{lem:fl}
Fix a picking sequence $s$ and $\succsim\in\cW^N$.
\begin{enumerate}[label=(\alph*),nosep]
\item For every $m\le T+1$ an $m$-partial assignment exists, $F_m\neq\emptyset$ for $m\le T$, and $F_{m'}\subseteq F_m$ for $m<m'\le T$; consequently $L_s(\succsim)\neq\emptyset$ and, for each agent $i$ with turns $m_1<\cdots<m_q$, $K_{m_1}\succsim_iK_{m_2}\succsim_i\cdots\succsim_iK_{m_q}$.
\item Let $\psi:\{1,\dots,T\}\to O$ be an injection and $\bar m\le T+1$ with $\psi(m)\in K_m$ for every $m<\bar m$. Then $\psi(m)\in C_m$ for every $m<\bar m$. In particular, if $\psi(m)\in K_m$ for every $m$, then $\psi$ generates an element of $L_s(\succsim)$.
\item $L_s$ satisfies \textsf{ESV}, \textsf{PI}, \textsf{SV} with $L_s(\succ)=\{\Dr_s(\succ)\}$ on $\cP^N$, and \NEU\ as a correspondence. Every selection from $L_s$ satisfies \NW. If $\Dr_s$ satisfies \EFO, every selection from $L_s$ satisfies \EFO\ with respect to $\succsim$, and if $\Dr_s$ satisfies \NEL, every selection satisfies \NEL\ with respect to $\succsim$.
\end{enumerate}
\end{lemma}

\begin{proof}
(a) If $T=0$, the empty assignment generates the unique empty allocation, and the remaining claims are vacuous. Suppose $T\ge1$. The empty map is a $1$-partial assignment and $F_1=O$. Inductively, suppose an $m$-partial assignment exists for some $m\le T$. It uses $m-1<T$ objects, so it avoids an object and $F_m\neq\emptyset$. Hence $C_m\neq\emptyset$. For any $z\in C_m\subseteq F_m$, choose an $m$-partial assignment avoiding $z$ and assign $z$ to turn $m$; this gives an $(m+1)$-partial assignment. The induction reaches $T+1$ without defining $F_{T+1}$, and a $(T+1)$-partial assignment generates an element of $L_s(\succsim)$. For $m<m'\le T$, the restriction of an $m'$-partial assignment to $\{1,\dots,m-1\}$ is an $m$-partial assignment avoiding the same objects, so $F_{m'}\subseteq F_m$. Thus the most preferred class of $F_{m'}$ is weakly below the most preferred class of $F_m$ for the same agent, which gives the ordering of her commitment classes.

(b) Suppose not, and let $m<\bar m$ be the smallest index with $\psi(m)\notin C_m$, i.e.\ $\psi(m)\in K_m\setminus F_m$. By minimality $\psi|_{\{1,\dots,m-1\}}$ is an $m$-partial assignment; since $\psi(m)\notin F_m$, every $m$-partial assignment uses $\psi(m)$, so $\psi(j)=\psi(m)$ for some $j<m$, contradicting injectivity.

(c) Any two elements of $L_s(\succsim)$ give agent $i$ one object of $K_m$ for each of her turns $m$, hence bundles with the same multiset of indifference classes, which dominate each other: \textsf{ESV}. If $y_i\sim_ix_i$ for all $i$, then by \autoref{lem:sorted} the $\ell$-th most preferred object of $y_i$ is indifferent to the $\ell$-th most preferred object of $x_i$, which by (a) lies in $K_{m_\ell}$; assigning the $\ell$-th most preferred object of $y_i$ to turn $m_\ell$ defines an injection $\psi$ with $\psi(m)\in K_m$, and (b) gives $y\in L_s(\succsim)$: \textsf{PI}. On a strict profile every $C_m$ is a singleton, $F_m=O\setminus\{\varphi(1),\dots,\varphi(m-1)\}$, and $C_m$ is the most preferred available object: \textsf{SV}. \NEU\ holds because the construction uses only $s$ and the reported orders. \NW\ is immediate. For \EFO\ and \NEL, fix a run $\varphi$ generating $x\in L_s(\succsim)$. At turn $m$ every object not used by $\varphi(1),\dots,\varphi(m-1)$ lies in $F_m$, so $\varphi(m)$ is $\succsim_{s_m}$-maximal among them, and the picker weakly prefers $\varphi(m)$ to every object assigned to a later turn. Suppose $\Dr_s$ satisfies \EFO. By \autoref{lem:sizes}, bundle sizes under $\Dr_s$ differ by at most one, and Step~1(b) in the proof of \autoref{thm:strict}(b), applied to $\Dr_s$ at a unanimous profile, shows that every agent's $\ell$-th turn comes before every other agent's $(\ell+1)$-th turn. For agents $i\neq j$, the claim is immediate if $x_j=\emptyset$; otherwise, mapping the object of $j$'s $(\ell+1)$-th turn to that of $i$'s $\ell$-th turn shows that $x_i$ dominates $x_j$ without the object of $j$'s first turn under $\succsim_i$, which is \EFO. Suppose instead that $\Dr_s$ satisfies \NEL, and let $|x_i|<|x_j|$. A set cannot dominate a larger set, so $|x_j|=|x_i|+1$, and \NEL\ at a unanimous profile, with \autoref{lem:sorted}, shows that $i$'s $t$-th turn comes before $j$'s $t$-th turn for every $t\le|x_i|$. Given $y\in x_j$ assigned to $j$'s $\ell$-th turn, map the objects of $j$'s $t$-th turns with $t<\ell$ to those of $i$'s $t$-th turns, and those with $t>\ell$ to those of $i$'s $(t-1)$-th turns. Each object of $j$ comes from a turn after the turn of its image, so $x_i$ dominates $x_j\setminus\{y\}$ under $\succsim_i$, which is \NEL.
\end{proof}

For $\succsim\in\cW^N$ let $\mathrm{Ref}(\succsim)$ be the set of strict profiles $\succ$ such that each $\succ_i$ \emph{refines} $\succsim_i$, that is, breaks its ties without changing its strict comparisons: $a\succ_ib$ implies $a\succsim_ib$.

\begin{lemma}\label{lem:nest}
Fix $\pi$ and $\succsim\in\cW^N$.
\begin{enumerate}[label=(\alph*),nosep]
\item The set $\{\Dr^\pi(\succ):\succ\in\mathrm{Ref}(\succsim)\}$ is the set of allocations obtained by running $s(\pi)$ with each picker taking, at each of her turns, an arbitrary object of her most preferred class among the available objects.
\item Every such allocation satisfies \NW, \EFO\ and \NEL\ with respect to $\succsim$, and every allocation in $L^\pi(\succsim)$ is one of them.
\end{enumerate}
\end{lemma}

\begin{proof}
(a) Under a refinement each pick is in the picker's most preferred available class. Conversely, let $y_1,\dots,y_T$ be the picks of such a run and let each agent $i$ refine $\succsim_i$ by ordering the objects within every indifference class so that the objects she picks come first, in the order she picks them. At each of her turns her pick is then the available object she ranks first under the refinement, since every other available object of its class is picked by her later or never. So the run is $\Dr^\pi$ at a refinement.

(b) Such an allocation is $\Dr^\pi(\succ)$ for a refinement $\succ$; \NW\ is immediate, and \EFO\ and \NEL\ transfer from \autoref{lem:draft} because an injection with $\mu(y)\succsim_iy$ under the refinement satisfies $\mu(y)\succsim_iy$ under $\succsim_i$. Every run of $L^\pi$ is of the form in (a) by the last step of the proof of \autoref{lem:fl}(c).
\end{proof}

\begin{lemma}\label{lem:topset}
Let $f$ satisfy \NW, \EFO\ and \WSP. Fix a profile $\succ$, an agent $i$ and a report $\succ'_i$, and suppose that $Z=f_i(\succ'_i,\succ_{-i})$ is the set of the $|Z|$ most preferred objects under $\succ_i$.
\begin{enumerate}[label=(\alph*),nosep]
\item If $|Z|=k+1$, or if $|Z|=k$ and $r=0$, then $f_i(\succ)=Z$.
\item If $|Z|=k$, then $f_i(\succ)=Z$ or $|f_i(\succ)|=k+1$.
\end{enumerate}
\end{lemma}

\begin{proof}
By \NW, \EFO\ and \autoref{lem:sizes}, every bundle has $k$ or $k+1$ objects at every profile, and $k+1$ only if $r\ge1$. Let $Y=f_i(\succ)\neq Z$ with $|Y|\le|Z|$. By \autoref{lem:sorted}, $Z\dom{i}Y$, because the $\ell$-th most preferred element of $Z$ is the $\ell$-th most preferred object overall; and $Y\not\dom{i}Z$, because a set cannot dominate a larger set and two distinct sets of the same size cannot mutually dominate each other under a strict order. Then $i$ gains by reporting $\succ'_i$, contradicting \WSP. In (a) every bundle has at most $|Z|$ objects, and in (b) the bundles with $k$ objects are excluded; this proves both parts.
\end{proof}

\section{Additional results}\label{app:add}

Appendices~\ref{app:neutral} and~\ref{app:cardinal} accompany the discussion of compensation in \autoref{sec:nel}, Appendix~\ref{app:weak} the treatment of indifferences in \autoref{sec:fl}, Appendix~\ref{app:mm} the characterizations of \autoref{sec:draft}, and Appendices~\ref{sec:separable} and~\ref{app:incentives} the incentive results of \autoref{sec:incentives}.

\subsection{The incomparability requirement}\label{app:neutral}

\autoref{sec:nel} compares \NEL\ with the incomparability requirement, a symmetric alternative. Like \AAU, it is stated at unanimous strict profiles, where differences between bundles reveal priority. A rule $f$ on $\cP^N$ satisfies the \emph{incomparability requirement} if, at every unanimous profile $\succ^u$, with $x=f(\succ^u)$, for all $i,j\in N$ with $|x_j|=|x_i|+1$, neither of the sets $x_i$ and $x_j\setminus\{x_j(|x_j|)\}$ dominates the other. In particular, two empty sets are not PD-incomparable, since each dominates the other. The next result shows what it selects among draft rules.

\begin{proposition}\label{prop:neutralstd}
Suppose that $r\ge1$.
\begin{enumerate}[label=(\alph*),nosep]
\item If $k\le1$, no non-wasteful rule $f:\cP^N\to\cA$ satisfies \EFO\ and the incomparability requirement.
\item If $k\ge2$, a draft rule $\Dr_s$ satisfies \EFO, \AAU\ and the incomparability requirement if and only if (i) rounds $1,\dots,k$ of $s$ each consist of $n$ picks, one by each agent, and the final round consists of $r$ picks by distinct agents; (ii) among the agents who receive $k$ objects, and among those who receive $k+1$, the order of picks reverses from one round to the next; and (iii) for every agent $i$ who receives $k$ objects and every agent $j$ who receives $k+1$, there are rounds $t,t'\le k$ such that $i$ picks before $j$ in round $t$ and after $j$ in round $t'$.
\end{enumerate}
\end{proposition}

\begin{proof}
Fix a unanimous profile $\succ^u$. By \autoref{lem:sizes}, non-wastefulness and \EFO\ at $\succ^u$ give $r$ agents $k+1$ objects and the others $k$ objects.

(a) Let $i$ receive $k$ objects and $j$ receive $k+1$; both exist because $1\le r\le n-1$. If $k=0$, the sets $x_i$ and $x_j\setminus\{x_j(1)\}$ are both empty, so each dominates the other. If $k=1$, they are single objects, one of which $\succ^u$ ranks above the other, so one set dominates the other. In either case the incomparability requirement fails.

(b) Let $\Dr_s$ satisfy \EFO, \AAU\ and the incomparability requirement. Draft rules are non-wasteful, so \autoref{lem:sizes} shows that $r$ agents receive $k+1$ objects and the others $k$, and Step~1(b) in the proof of \autoref{thm:strict}(b) shows that every agent's $t$-th pick comes before every other agent's $(t+1)$-th pick, which is (i). At $\succ^u$ the $m$-th picker receives the $m$-th ranked object, so $x_i(t)$ is agent $i$'s pick in round $t$, and $x_i(t)\succ^u x_j(t)$ holds exactly when $i$ picks before $j$ in round $t$. \AAU\ then gives (ii), as in Step~1(d) of that proof. For $i$ with $k$ objects and $j$ with $k+1$, \autoref{lem:sorted} shows that $x_i$ dominates $x_j\setminus\{x_j(k+1)\}$ if and only if $i$ picks before $j$ in every round $t\le k$, and that $x_j\setminus\{x_j(k+1)\}$ dominates $x_i$ if and only if $j$ picks before $i$ in every such round. The incomparability requirement therefore holds exactly when (iii) does.

Conversely, let $s$ satisfy (i)--(iii). By (i), every bundle has $k$ or $k+1$ objects, and for agents $i\neq j$ and every $\ell$, $j$'s $\ell$-th pick comes before $i$'s $(\ell+1)$-th pick, so that $j$ prefers it by \autoref{lem:draft}(a). Mapping $i$'s $(\ell+1)$-th pick to $j$'s $\ell$-th pick shows that, at every profile, $x_j$ dominates $x_i$ without $i$'s first pick, which is \EFO. At a unanimous profile, (ii) gives \AAU\ and (iii) gives the incomparability requirement, by the equivalences above.
\end{proof}

The family in (b) contains the plain snake, which reverses the order of all agents from one round to the next, but not the block--snake draft, in which the agents with fewer objects pick first in every round. Unlike \NEL, the incomparability requirement does not determine the order up to the draw. With two agents and $2k+1$ objects, the first picker under the plain snake receives the extra object exactly when $k$ is even. The broader family admits orders that start with either the agent receiving fewer objects or the agent receiving more. Part (a) reflects the formalization rather than the idea behind the incomparability requirement. When the agents with fewer objects receive at most one object, the sets compared have at most one element, and two such sets are always comparable.

\subsection{A cardinal comparison}\label{app:cardinal}

\autoref{sec:nel} notes that \NEL, being ordinal, can overcompensate under cardinal benchmarks. \autoref{tab:cardinal} reports how large the gap between the two groups is on average under four models, and \autoref{tab:realized} how it varies across realizations under the two stochastic models. Agents pick sincerely, and an agent's utility is the sum of her values for her objects, divided by her fair share, her value for all objects divided by $n$. Under \emph{Borda}, all agents share the values $T,T-1,\dots,1$. Under \emph{steep}, they share the value $0.7^{m-1}$ for the $m$-th ranked object. Under \emph{uniform private}, each agent's values are independent draws from the uniform distribution on $[0,1]$, and under \emph{correlated}, each value is the average of a draw common to all agents and a private draw. Entries for the random models average 100{,}000 draws (seed 2026), with standard errors of at most $0.11$ points, and all orders in a problem and model are evaluated on the same draws; both tables use the same draws. Under the plain snake and round-robin, the agents who receive $k+1$ objects are those who pick in the last, incomplete round.

\begin{table}[ht]
\centering\small
\begin{tabular}{llrrrr}
\toprule
Problem & Order & Borda & Steep & Uniform private & Correlated\\
\midrule
$n=2$, $T=19$ & Block--snake & $8.4$ & $35.2$ & $1.8$ & $1.1$\\
 & Plain snake & $0.0$ & $12.0$ & $-2.5$ & $-3.1$\\
 & Thue--Morse & $0.0$ & $-6.6$ & $-2.4$ & $-3.5$\\
 & Round-robin & $-10.5$ & $-35.4$ & $-8.3$ & $-9.5$\\
\midrule
$n=3$, $T=7$ & Block--snake & $21.4$ & $35.9$ & $-9.5$ & $-6.9$\\
 & Plain snake & $-10.7$ & $-39.1$ & $-20.7$ & $-23.7$\\
 & Round-robin & $-42.9$ & $-64.9$ & $-41.3$ & $-42.4$\\
\midrule
$n=3$, $T=19$ & Block--snake & $12.6$ & $49.1$ & $0.5$ & $1.0$\\
 & Plain snake & $-1.6$ & $-28.8$ & $-5.4$ & $-6.5$\\
 & Round-robin & $-15.8$ & $-55.6$ & $-14.4$ & $-15.3$\\
\midrule
$n=4$, $T=19$ & Block--snake & $12.6$ & $76.8$ & $-12.6$ & $-6.6$\\
 & Plain snake & $-4.2$ & $-34.5$ & $-17.4$ & $-15.9$\\
 & Round-robin & $-21.1$ & $-61.3$ & $-23.7$ & $-23.0$\\
\bottomrule
\end{tabular}
\caption{Mean normalized utility of the agents who receive $k$ objects minus that of the agents who receive $k+1$, in percentage points of a fair share. Positive entries mean that those without the windfall come out ahead.}\label{tab:cardinal}
\end{table}

\begin{table}[ht]
\centering\small
\begin{tabular}{llrrrr}
\toprule
 & & \multicolumn{2}{c}{Uniform private} & \multicolumn{2}{c}{Correlated}\\
\cmidrule(lr){3-4}\cmidrule(lr){5-6}
Problem & Order & Mean $|$gap$|$ & Ahead (\%) & Mean $|$gap$|$ & Ahead (\%)\\
\midrule
$n=2$, $T=19$ & Block--snake & $10.5$ & $55.6$ & $6.0$ & $55.3$\\
 & Plain snake & $10.6$ & $42.5$ & $6.5$ & $33.7$\\
 & Thue--Morse & $10.6$ & $42.8$ & $6.7$ & $31.8$\\
 & Round-robin & $12.4$ & $25.7$ & $10.2$ & $9.5$\\
\midrule
$n=3$, $T=7$ & Block--snake & $27.0$ & $39.5$ & $17.0$ & $36.2$\\
 & Plain snake & $30.2$ & $26.0$ & $25.8$ & $11.2$\\
 & Round-robin & $42.9$ & $6.9$ & $42.5$ & $0.4$\\
\midrule
$n=3$, $T=19$ & Block--snake & $13.2$ & $52.6$ & $7.3$ & $55.1$\\
 & Plain snake & $13.7$ & $38.2$ & $9.0$ & $23.8$\\
 & Round-robin & $17.4$ & $18.4$ & $15.5$ & $3.1$\\
\midrule
$n=4$, $T=19$ & Block--snake & $17.6$ & $22.2$ & $9.1$ & $22.2$\\
 & Plain snake & $20.6$ & $15.2$ & $16.3$ & $4.8$\\
 & Round-robin & $25.4$ & $8.4$ & $23.1$ & $0.7$\\
\bottomrule
\end{tabular}
\caption{Variation across realizations in the two stochastic models, on the draws of \autoref{tab:cardinal}. Mean $|$gap$|$ is the mean absolute value of the gap between the groups' mean normalized utilities, in percentage points of a fair share, and Ahead is the share of draws in which the agents who receive $k$ objects come out ahead of those who receive $k+1$. Standard errors are at most $0.11$ points for the means and $0.2$ percentage points for the shares.}\label{tab:realized}
\end{table}

In the common-value models and parameter cases reported here, the block--snake draft overcompensates, more so when values fall steeply, and except under steep values with $n=3$ and $T=7$, the plain snake or the Thue--Morse order gives a smaller absolute gap between the groups. In the two stochastic models, the block--snake draft has the lowest estimated mean absolute gap in every reported problem (\autoref{tab:realized}). In the two-agent uniform model its margin over the plain snake and the Thue--Morse order is about $0.1$ points, several standard errors of the paired difference on the common draws but economically small; elsewhere the margin is at least $0.5$ points. The share of draws in which the agents with fewer objects come out ahead is close to one half under the block--snake draft when objects per agent are many, but in the two problems with few objects per agent ($n=3$, $T=7$ and $n=4$, $T=19$) the group with the windfall is ahead in 60 to 78 percent of draws even under the block--snake draft, and in more under the plain snake and round-robin. These are comparisons of group means in the stated benchmarks, not of every pair of agents; heterogeneous preferences alone do not guarantee a small gap or the same ranking of the rules.

\subsection{The weak preference domain}\label{app:weak}

This subsection explains why a fair, neutral rule must be set-valued and records what is known toward characterizing the forward-looking draft.

\begin{proposition}\label{prop:noneutral}
Let $n\ge2$ and $T\ge2$. No single-valued rule $g:\cW^N\to\cA$ satisfies \NW, \EFO\ and \NEU.
\end{proposition}

\begin{proof}
Let $I$ be the profile at which every agent is indifferent among all objects. For every bijection $\sigma$ of $O$ we have $I^\sigma=I$, so \NEU\ gives $\sigma(g(I))=g(I)$; hence each bundle $g_i(I)$ is invariant under all permutations of $O$, i.e.\ $g_i(I)\in\{\emptyset,O\}$. By \NW\ some agent $j$ receives $O$ and every other agent receives $\emptyset$. For $i\neq j$, \EFO\ requires $\emptyset\dom{i}O\setminus\{y\}$ for some $y$, which forces $T\le1$.
\end{proof}

Every allocation of the forward-looking draft arises from some way of breaking the ties, but not every way of breaking them yields one (\autoref{lem:nest} in \autoref{app:aux} and \autoref{ex:three}). We do not have a characterization of $L^\pi$ on the full domain of weak orders on objects. What distinguishes it from other PD-efficient ways of resolving ties is a serial priority in tie resolution, and we do not know a condition stated solely in terms of allocations and preferences that characterizes this tie resolution.

The forward-looking draft can be computed by bipartite matching. Whether an object $z$ lies in $F_m$ is the question whether turns $1,\dots,m-1$ can be matched to distinct objects of their commitment sets other than $z$, a bipartite matching problem. It suffices to keep one matching of turns $1,\dots,m-1$ to their commitment sets and to extend it by a single augmenting path when turn $m$ is added. An object can then be avoided if and only if it is unmatched or reachable from an unmatched object by an alternating path, so one search finds all of $F_m$. Each turn takes $O(T^2)$ time, the whole draft $O(T^3)$, and the final matching is an allocation.

\subsection{Object-wise Maskin monotonicity and the independence of the conditions}\label{app:mm}

Object-wise Maskin monotonicity separates into an informational part and a monotonicity part. At a unanimous profile a rule's outcome reveals the order in which agents are served, and the condition says that this order can depend neither on how an agent ranks objects she does not receive nor on changes that only make her like her own objects more.

\begin{description}[leftmargin=1.6em,style=nextline]
\item[\IUC\ (independence of unreceived comparisons)] Let $x=f(\succ)$, $i\in N$, and let $\succ'_i$ agree with $\succ_i$ on every pair of objects at least one of which lies in $x_i$. Then $f(\succ'_i,\succ_{-i})=x$.
\item[\ORa\ (raising a received object)] Let $x=f(\succ)$, $i\in N$, $y\in x_i$, and let $\succ'_i$ be obtained from $\succ_i$ by moving $y$ to a higher position, passing only objects not in $x_i$. Then $f(\succ'_i,\succ_{-i})=x$.
\end{description}
\IUC\ says that the rule never uses an agent's comparison between two objects she does not receive; \ORa\ says that an agent who comes to like a received object more, relative to objects she does not receive, keeps her bundle.

\begin{lemma}\label{lem:decomp}
A rule $f$ satisfies \MM\ if and only if it satisfies \IUC\ and \ORa.
\end{lemma}

\begin{proof}
Throughout, fix $\succ$, $i\in N$ and $x=f(\succ)$, and call $\succ'_i$ an \emph{\MM-move for $i$} if $\{z:y\succ_i z\}\subseteq\{z:y\succ'_i z\}$ for every $y\in x_i$. Since the hypothesis of \MM\ is a conjunction of conditions on individual agents, \MM\ is equivalent to invariance under single-agent \MM-moves: if each such move leaves the allocation at $x$, the agents' moves can be made one at a time, each evaluated at the same allocation $x$.

(\MM\ $\Rightarrow$ \IUC, \ORa.) Let $\succ'_i$ be as in \IUC. For $y\in x_i$, every pair $\{y,z\}$ is ranked the same way at $\succ_i$ and $\succ'_i$, so the lower contour set of $y$ is unchanged; $\succ'_i$ is an \MM-move. Let $\succ'_i$ be as in \ORa, obtained by moving $y\in x_i$ upward past objects outside $x_i$. The lower contour set of $y$ gains the objects passed; for $y'\in x_i\setminus\{y\}$, neither the position of $y$ relative to $y'$ nor that of any passed object relative to $y'$ changes, so the lower contour set of $y'$ is unchanged. Again $\succ'_i$ is an \MM-move.

(\IUC, \ORa\ $\Rightarrow$ \MM.) Let $\succ'_i$ be an \MM-move for $i$. First, objects of $x_i$ keep their relative order: if $y\succ_i y'$ and $y'\succ'_i y$ for $y,y'\in x_i$, then $y'$ leaves the lower contour set of $y$. Second, an object $z\notin x_i$ that lies below some $y\in x_i$ at $\succ_i$ lies below $y$ at $\succ'_i$; thus objects outside $x_i$ cross objects of $x_i$ only downward. Write $x_i=\{y^1,\dots,y^q\}$ with $y^1\succ_i\cdots\succ_i y^q$. We transform $\succ_i$ into $\succ'_i$ by a chain of \IUC- and \ORa-moves, each evaluated at the allocation $x$, which both conditions leave unchanged. For $j=1,\dots,q$ in turn: let $Z_j$ be the set of objects currently above $y^j$ that lie below $y^j$ at $\succ'_i$. Each $z\in Z_j$ lies below $y^{j-1}$ at the current order (if $j\ge2$): otherwise $z$ would be above $y^{j-1}$, and after the previous step the objects above $y^{j-1}$ are exactly those $z'$ with $z'\succ'_i y^{j-1}$, giving $z\succ'_i y^{j-1}\succ'_i y^j$, a contradiction. Hence $Z_j$ lies in the segment between $y^{j-1}$ and $y^j$ (or above $y^1$ if $j=1$), which contains no object of $x_i$. Permute that segment so that $Z_j$ sits immediately above $y^j$ (an \IUC-move: only pairs of objects outside $x_i$ change), and then move $y^j$ upward past $Z_j$ (an \ORa-move). After this step the objects above $y^j$ are exactly those $z$ with $z\succ'_i y^j$. After step $q$ the current order agrees with $\succ'_i$ on every pair involving an object of $x_i$; a final \IUC-move permutes the objects outside $x_i$ within each segment to match $\succ'_i$.
\end{proof}

Both conditions are satisfied by every draft rule, and \ORa\ is the substantive one: with $n=2$, $k=2$ and $r=0$, the rule in the proof of \autoref{prop:indep} that violates \MM\ satisfies \IUC\ and violates \ORa.

\begin{proposition}\label{prop:indep}
For each of \NW, \NEL, \AAU, \NEU\ and \MM, under the parameter restrictions indicated in the proof, some rule satisfies the other five conditions of \autoref{thm:strict}(b) but not this one. For \EFO, see \autoref{cor:ef1}.
\end{proposition}

The construction for \NEU\ requires $n\ge3$. With $n=2$ and $T=2$, independence is immediate: for $O=\{a,b\}$, the constant rule $f(\succ)=(\{a\},\{b\})$ satisfies the other five conditions but violates \NEU. The remaining two-agent parameter cases are not resolved by these constructions.

\begin{proof}
The rules used below for \NEU\ and \MM\ select, at each profile, the outcome of some block--snake draft $\Dr^{\pi'}$, so \NW, \EFO, \NEL\ and \AAU\ hold for them automatically, since each of these conditions constrains the allocation at a single profile and is satisfied by every $\Dr^{\pi'}$ (\autoref{lem:draft}).

\emph{\NW\ (any $T\ge1$).} The null rule $f\equiv(\emptyset,\dots,\emptyset)$ satisfies \EFO\ and \NEL\ (all bundles are empty), \AAU\ (vacuously), \NEU\ ($\sigma(\emptyset)=\emptyset$) and \MM\ (it is constant), and violates \NW.

\emph{\NEL\ ($k\ge1$ and $r\ge1$; when $k=0$, every rule satisfying \NW\ and \EFO\ gives at most one object to each agent, and \NEL\ holds automatically).} The \emph{$B$-first draft}: the picking sequence
\[
B^1 A^1\;B^2 A^2\;\cdots\;B^k A^k\;B^{k+1} .
\]
It is a draft rule, hence satisfies \NW, \NEU\ and \MM\ (the proofs in \autoref{lem:draft} do not use the order of the blocks); it has round structure with $B$'s extra picks last, hence \EFO; and each block is reversed every round, hence \AAU. It violates \NEL: at a unanimous profile with common order $z_1\succ^u\cdots\succ^u z_T$, the first $B$ agent holds $z_1$ while every $A$ agent's first object has rank larger than $r$, so after removing the $B$ agent's last object her $t$-th most preferred object is still ranked above the $A$ agent's $t$-th most preferred object for every $t$.

\emph{\AAU\ ($k\ge2$ and $n-r\ge2$, or $k\ge1$ and $r\ge2$).} \emph{Round-robin within blocks}: the sequence
\[
A^1 B^1\;A^1 B^1\;\cdots\;A^1 B^1\;B^1
\]
(no reversal). It is a draft rule whose sequence has round structure, with the $A$-block first in every round, so \NW, \EFO, \NEL, \NEU, \MM\ hold; two agents of the same block keep the same precedence in consecutive rounds, so \AAU\ fails.

\emph{\NEU\ ($n\ge3$ and $k\ge1$).} Fix $a\in O$ and two bijections $\pi_1\neq\pi_2$ with $\pi_1(1)=\pi_2(1)=1$. Let $f(\succ)=\Dr^{\pi_1}(\succ)$ if $a$ is agent $1$'s most preferred object and $f(\succ)=\Dr^{\pi_2}(\succ)$ otherwise. \MM: agent $1$ picks first under either draw, so her most preferred object $y$ lies in $x_1$; an \MM-improvement keeps every object below $y$ below $y$, hence $y$ remains her most preferred object, the trigger is unchanged, and \MM\ of $\Dr^{\pi_1}$ or $\Dr^{\pi_2}$ gives $f(\succ')=x$. \NEU\ fails: take a unanimous profile $\succ$ with $a$ on top and a bijection $\sigma$ swapping $a$ with another object; then $f(\succ^\sigma)=\Dr^{\pi_2}(\succ^\sigma)=\sigma(\Dr^{\pi_2}(\succ))\neq\sigma(\Dr^{\pi_1}(\succ))=\sigma(f(\succ))$, because $\pi_1\neq\pi_2$ yield different allocations at unanimous profiles.

\emph{\MM\ ($k\ge2$, or $k=1$ and $r\ge2$).} Choose draws $\pi_1,\pi_2$ as follows. If $k\ge2$ and $n-r\ge2$, or if $r\ge2$, let $\pi_1$ list agents $1$ and $2$ consecutively, first in the block $A(\pi_1)$ in the first case and first in $B(\pi_1)$ otherwise, and let $\pi_2$ be $\pi_1$ with agents $1$ and $2$ exchanged. In the remaining case, $n=2$ and $r=1$, let $\pi_1=(1,2)$ and $\pi_2=(2,1)$. Let $f(\succ)=\Dr^{\pi_1}(\succ)$ if agents $1$ and $2$ have the same most preferred object and $f(\succ)=\Dr^{\pi_2}(\succ)$ otherwise. \NEU\ holds because relabeling preserves ``same most preferred object'', and at unanimous profiles $f=\Dr^{\pi_1}$. To see that \MM\ fails, fix distinct objects $a,b,c$; let agent $1$ rank $a\succ b\succ c$ above all other objects, agent $2$ rank $a\succ c\succ b$ above all other objects, and every other agent rank $a$, $b$ and $c$ below all other objects. The tops coincide, so $x=f(\succ)=\Dr^{\pi_1}(\succ)$, and agent $2$, who picks right after agent $1$, receives $c$ in round $1$. Let $\succ'$ differ from $\succ$ only in that agent $2$ moves $c$ to the top. This raises the received object $c$ past the unreceived object $a$ only, so it is an \MM-improvement of $x$; but the tops now differ, and $f(\succ')=\Dr^{\pi_2}(\succ')$. In the first two cases, agents $1$ and $2$ still take $a$ and $c$ in round $1$, but in round $2$, where the order of their block is reversed, agent $1$ now picks before agent $2$ and takes $b$, which $x$ gives to agent $2$; the other agents never take $a$, $b$ or $c$ before these picks, since enough other objects remain. In the third case agent $2$ moves from the block of size $k+1$ to the block of size $k$. Either way $f(\succ')\neq x$.
\end{proof}

\subsection{Separate types}\label{sec:separable}\label{app:separable}

Suppose that the objects come in types, such as games grouped by month, and that fairness is assessed type by type. A full type contains $n$ objects, one for each agent; a short type, when present, contains $r$ objects with $1\le r<n$. Running a serial dictatorship within each type, in the block--snake priority pattern, is then strategy-proof, because a report in one type affects neither the allocation nor the available choices in any other \citep{Caspari2023}. Without a short type, this is the booster draft of \citet{Caspari2026}, who proposes it as a strategy-proof alternative to the draft and shows that it satisfies his half-bundle envy bound; our typewise principles give its reversal of priorities a normative reading, as alternation, and extend the construction to a short type (\autoref{prop:separable}).

\begin{remark}\label{rem:slotwise}
Type-by-type fairness never compares objects of different types. It is the natural standard when the institution fixes the types, as when each fan is to attend one game per month, but not when agents care only about their total bundles. If agents rank all objects across types and compare bundles by pairwise dominance, as in \autoref{sec:model}, the block--snake pattern implemented within types can violate both \EFO\ and \NEL, and neither the bundle-level nor the type-by-type compensation principle implies the other (see the examples below). Comparing the two environments under a single fairness standard is left for future work.
\end{remark}

Formally, the object set is partitioned into \emph{types} $O=O_1\cup\cdots\cup O_{k+1}$ with $|O_t|=n$ for $t\le k$ and $|O_{k+1}|=r\ge1$ (games grouped by month). Types are ordered, with the short type last. The order is a design choice, and the type-by-type principles below refer to it. We assume $k\ge1$, since with $k=0$ there is no full type and the principles below are vacuous. Let $\bot$ denote a dummy object standing for ``no object of the short type.'' The dummy is reusable and is excluded from the restriction that each real object be assigned at most once. A \emph{typewise allocation} assigns to every agent one object $x_i^t\in O_t$ of each full type $t\le k$ and one element $x_i^{k+1}\in O_{k+1}\cup\{\bot\}$, with each real object assigned to at most one agent and every object of $O_{k+1}$ assigned; exactly $r$ agents thus receive a short-type object. Each agent $i$ reports a strict ranking $\succ_i^t$ on $O_t$ for every $t\le k$ and a strict ranking $\succ_i^{k+1}$ on $O_{k+1}\cup\{\bot\}$ in which $\bot$ is last. We compare bundles by \emph{typewise dominance}: one bundle dominates another if its object is at least as good in every type,
\[
x_i\succsim^{\mathrm{tw}}_iy_i\quad\Longleftrightarrow\quad x_i^t\succsim_i^ty_i^t\ \text{ for every }t\in\{1,\dots,k+1\}.
\]
A \emph{rule} maps profiles of type rankings to typewise allocations; it is \emph{strategy-proof with respect to typewise dominance} if $f_i(\succ)\succsim^{\mathrm{tw}}_if_i(\succ'_i,\succ_{-i})$ for all $i$, $\succ$ and $\succ'_i$.

A \emph{typewise serial dictatorship} is determined by a profile $(\tau_1,\dots,\tau_{k+1})$ of linear orders on $N$: in every full type $t$, agents choose their favorite remaining object of $O_t$ in the order $\tau_t$. In the short type, agents choose in the order $\tau_{k+1}$, taking their favorite remaining real object if one is available and otherwise receiving $\bot$; the first $r$ agents of $\tau_{k+1}$ therefore receive objects and the others receive $\bot$. For categories of equal size, \citet{Caspari2023} characterizes these rules by strategy-proofness, category-wise neutrality (invariance to relabeling objects within a type) and nonbossiness (his categories are our types); our result is stated within the class and does not depend on extending that characterization to a short category.

The fairness principles are read type by type. Write $A$ for the agents who receive $k$ objects and $B$ for those who receive $k+1$; at a typewise serial dictatorship, $B$ is the set of the first $r$ agents of $\tau_{k+1}$.
\begin{description}[leftmargin=1.6em,style=nextline]
\item[\NEL$^{\mathrm{tw}}$ (no envy toward larger bundles, by type)] For all $i\in A$, $j\in B$ and every full type $t$: $x_i^t\succsim_i^tx_j^t$.
\item[\AAU$^{\mathrm{tw}}$ (alternating advantage by type)] At every profile in which all agents report the same ranking on every type, write $\succ^{u,t}$ for the common ranking in type $t$. For all $i,j$ with the same number of objects and every $t$ such that both receive objects of types $t$ and $t+1$ (receiving $\bot$ is not receiving an object), if $x_i^t\succ^{u,t}x_j^t$ then $x_j^{t+1}\succsim^{u,t+1}x_i^{t+1}$.
\end{description}
\NEL$^{\mathrm{tw}}$ is the compensation principle read type by type; \autoref{rem:slotwise} discusses when this typewise standard is appropriate and how it relates to the bundle-level \NEL. \EFO\ has no role, since ``one object per type'' is built into the model.

\begin{proposition}\label{prop:separable}
Let $f$ be a typewise serial dictatorship with orders $(\tau_1,\dots,\tau_{k+1})$, let $B$ be the set of the first $r$ agents of $\tau_{k+1}$ and $A=N\setminus B$.
\begin{enumerate}[label=(\alph*),nosep]
\item $f$ satisfies \NEL$^{\mathrm{tw}}$ if and only if, in every full type $t$, every agent of $A$ precedes every agent of $B$ in $\tau_t$. The orders within $A$ and within $B$ are unrestricted.
\item $f$ satisfies \NEL$^{\mathrm{tw}}$ and \AAU$^{\mathrm{tw}}$ if and only if, in addition, the order of $A$ in $\tau_{t+1}$ is the reverse of its order in $\tau_t$ for $t<k$, and the order of $B$ in $\tau_{t+1}$ is the reverse of its order in $\tau_t$ for $t\le k$. These rules implement the block--snake priority pattern of \autoref{sec:draft}, $A^1B^1,\ A^2B^2,\dots$, independently within each type, for the draw $\pi$ listing $A$ in the order of $\tau_1$ followed by $B$ in the order of $\tau_1$.
\item Every typewise serial dictatorship is strategy-proof; in particular so are all rules in (a) and (b).
\end{enumerate}
\end{proposition}

\begin{proof}
(a) If every agent of $A$ precedes every agent of $B$ in $\tau_t$, then at any profile an agent $i\in A$ chooses in type $t$ while the object $x_j^t$ of any $j\in B$ is still available, so $x_i^t\succsim_i^tx_j^t$. Conversely, suppose some $j\in B$ precedes some $i\in A$ in $\tau_t$, and take a profile in which all agents report the same ranking on $O_t$; the objects of type $t$ are then taken in the order $\tau_t$, so $x_j^t\succ_i^tx_i^t$ and \NEL$^{\mathrm{tw}}$ fails.

(b) Fix a profile in which all agents report the same ranking on every type, and write $\succ^{u,t}$ for the common ranking in type $t$; in each type the objects are taken in the order of that type's priority. For two agents both in $A$ or both in $B$, hence of the same size, $x_i^t\succ^{u,t}x_j^t$ means that $i$ precedes $j$ in $\tau_t$, and $x_j^{t+1}\succsim^{u,t+1}x_i^{t+1}$, which is strict since the objects are distinct, means that $j$ precedes $i$ in $\tau_{t+1}$. Hence \AAU$^{\mathrm{tw}}$ holds if and only if, for every pair in $A$ and every $t<k$, and for every pair in $B$ and every $t\le k$, precedence is reversed between $\tau_t$ and $\tau_{t+1}$, i.e.\ the restricted orders are reversed. Together with (a) this is the block--snake structure.

(c) This is the standard argument for serial dictatorship, applied type by type: a report in one type affects neither the allocation nor the available choices in any other.
\end{proof}

The following examples support \autoref{rem:slotwise}. Suppose that agents rank all objects across types and compare bundles by pairwise dominance.

For \NEL, take one full type $\{p,q\}$, a short type $\{s\}$ and two agents, with $i\in A$ choosing first in the full type. If $i$ ranks $s\succ p\succ q$ and $j$ ranks $p\succ q\succ s$, then $i$ receives $\{p\}$ and $j$ receives $\{q,s\}$, and after $q$ is removed $i$ strictly prefers $\{s\}$ to her own bundle.

For \EFO, take two full types $\{a_1,b_1\}$ and $\{a_2,b_2\}$, a short type $\{s\}$ and two agents, with $i\in A$, so that $i$ chooses first in both full types, and let both agents rank $a_1\succ a_2\succ b_1\succ b_2\succ s$. Then $i$ receives $\{a_1,a_2\}$ and $j$ receives $\{b_1,b_2,s\}$, and whichever object is removed from $i$'s bundle, the remaining one is ranked by $j$ above everything she holds.

Conversely, take full types $\{p,q\}$ and $\{u,v\}$, a short type $\{s\}$, and two agents with $x_i=\{q,u\}$ and $x_j=\{p,v,s\}$. If $i$ ranks $u\succ p\succ q\succ v\succ s$, she ranks her type-$1$ object below $j$'s, so \NEL$^{\mathrm{tw}}$ fails, yet her bundle dominates $j$'s bundle with any one object removed, so the bundle-level \NEL\ holds.

\subsection{Incentives}\label{app:incentives}

\begin{lemma}\label{lem:manip}
Let $s$ be a picking sequence in which some agent $i$ has two turns $p<q$ with no turn of $i$ and at least one turn of another agent strictly between them. Then $\Dr_s$ is not weakly strategy-proof.
\end{lemma}

\begin{proof}
Let $d_1,\dots,d_{p-1}$, $a$, $b$, $c_1,\dots,c_{q-p-1}$ be distinct objects and let $R$ be the remaining objects in some fixed order. Every agent other than $i$ reports $d_1\succ\cdots\succ d_{p-1}\succ b\succ c_1\succ\cdots\succ c_{q-p-1}\succ R\succ a$, and $i$ truthfully reports $d_1\succ\cdots\succ d_{p-1}\succ a\succ b\succ c_1\succ\cdots\succ c_{q-p-1}\succ R$. The first $p-1$ picks are $d_1,\dots,d_{p-1}$ regardless of who makes them. Truthfully, $i$ takes $a$ at turn $p$, the other agents take $b,c_1,\dots,c_{q-p-2}$ at turns $p+1,\dots,q-1$, and $i$ takes $c_{q-p-1}$ at turn $q$. If $i$ instead reports $b$ above $a$, she takes $b$ at turn $p$, the other agents take $c_1,\dots,c_{q-p-1}$, and $i$ takes $a$ at turn $q$: at each intervening turn a $c$-object remains available and is preferred to $a$ by the picker. After turn $q$ the sets of remaining objects coincide and $i$'s report ranks them identically, so her later picks are the same. Her bundle under the misreport replaces $c_{q-p-1}$ by $b$, which she strictly prefers, so it strictly dominates her truthful bundle.
\end{proof}

\citet{TroyanMorrill2020} call a manipulation \emph{obvious} if it yields a higher utility than truth-telling in the worst case or in the best case over the others' reports. We evaluate outcomes by an additive utility with positive values consistent with the agent's true ranking.

\begin{proposition}\label{prop:nom}
In every draft rule, and so in the block--snake draft, truth-telling maximizes both an agent's worst-case and her best-case utility over the others' reports, for every additive utility with positive values consistent with her ranking. Hence every draft rule is maxmin strategy-proof and not obviously manipulable.
\end{proposition}

\citet{CB2026} prove both properties for round-robin draft rules, and the argument for every picking sequence is in \autoref{app:proofs}.

For any number of agents, nonbossiness together with full strategy-proofness extends the conflict beyond picking orders. A rule is \emph{nonbossy} if no agent can change anyone else's bundle by changing her report without also changing her own:
\begin{description}[leftmargin=1.6em,style=nextline]
\item[Nonbossiness] For all $\succ$, $i$ and $\succ'_i$: if $f_i(\succ'_i,\succ_{-i})=f_i(\succ)$, then $f(\succ'_i,\succ_{-i})=f(\succ)$.
\end{description}

\begin{proposition}\label{prop:spnb}
\begin{enumerate}[label=(\alph*),nosep]
\item Every strategy-proof and nonbossy rule $f:\cP^N\to\cA$ satisfies \MM.
\item A rule is non-wasteful, neutral, strategy-proof and nonbossy if and only if it is a quota serial dictatorship.
\end{enumerate}
\end{proposition}

Part (a) connects \MM, the invariance condition of \autoref{thm:drafts}, to two familiar incentive properties, stated for object rankings and pairwise dominance. Part (b) follows by combining part (a) with \autoref{thm:drafts} and \autoref{lem:manip}; it characterizes quota serial dictatorships for reported object rankings, with non-wastefulness in place of the efficiency used in the characterization of \citet{Hatfield2009}. Since quota serial dictatorships violate \EFO\ whenever at least two agents receive more than one object (\autoref{prop:nelsp}(b)), part (b) has an immediate consequence:

\begin{corollary}\label{cor:nb}
If at least two agents receive more than one object, no non-wasteful, neutral, strategy-proof and nonbossy rule $f:\cP^N\to\cA$ satisfies \EFO.
\end{corollary}

\autoref{cor:nb} holds for any numbers of agents and objects but requires nonbossiness and full strategy-proofness; for two agents, \autoref{prop:imposs} dispenses with both.

\begin{remark}\label{rem:kone}
Among draft rules, \EFO\ and weak strategy-proofness are incompatible whenever at least two agents receive more than one object (\autoref{prop:nelsp}(b)), but among all rules this condition does not suffice. Let $n\ge3$, $k=1$ and $r\ge2$, so that $r$ agents receive two objects, and fix an agent $a$ who receives one object and an order of the $r$ agents who receive two. Agent $a$ receives the object she ranks first. Her ranking then splits the remaining objects into a first pool, her next $n-1$ objects, and a second pool, her last $r$ objects. In the first pool the other agents who receive one object choose first, in a fixed order, followed by the agents who receive two objects in their fixed order, and in the second pool the latter choose in the reverse order. The rule is non-wasteful and neutral, since the pools are defined by a ranking and not by labels. It is strategy-proof, since truth-telling gives $a$ her most preferred object whatever the others report, and every other agent's choice in one pool affects neither the allocation nor her available choices in the other. It satisfies \EFO, since each agent who receives one object either holds her most preferred object or chooses in the first pool before every agent who receives two, and any two agents who receive two objects choose in opposite orders in the two pools. It is bossy, because $a$'s ranking determines the pools, so nonbossiness cannot be dropped from \autoref{cor:nb}. \autoref{prop:imposs} accordingly assumes that every agent receives at least two objects.
\end{remark}

\citet{CB2026} show that no rule that respects a priority satisfies \EFO, non-wastefulness and weak strategy-proofness (their Theorem~2), and that no PD-efficient rule satisfies \EFO\ and weak strategy-proofness (their Theorem~3); an earlier version shows that, with two agents, strategy-proofness is incompatible with \EFO\ and non-wastefulness in their model, where the set of objects varies \citep{CB2024}. When every agent receives the same number of objects, at least two, \citet[Theorem~1]{Caspari2026} gives an impossibility for any number of agents, since no non-wasteful and PD-efficient rule that satisfies \EFO\ is then weakly strategy-proof.\footnote{\label{fn:caspari-bound}With $r=0$, non-wastefulness and \EFO\ give every agent $k$ objects. For agents $i$ and $j$, write $x_i(\ell)$ and $x_j(\ell)$ for the $\ell$-th most preferred objects of $x_i$ and $x_j$ under $\succ_i$. \EFO\ for agents $i$ and $j$ then means that $x_i(\ell)\succsim_ix_j(\ell+1)$ for every $\ell\le k-1$, and hence $x_i(\ell)\succsim_ix_j(\ell+d)$ for every $d\ge1$ and $\ell\le k-d$. Taking $d=\lceil k/2\rceil$ gives Caspari's $\lceil k/2\rceil$-envy-freeness, under which envy disappears once the $d$ most preferred objects of $x_j$ and the $d$ least preferred objects of $x_i$ are removed. PD-efficiency implies his Pareto possibility, and on bundles of equal size his dominance relation is pairwise dominance.}

\section{Proofs of the results in the text}\label{app:proofs}

\begin{proof}[Proof of \autoref{cor:ef1}]
Let $\Dr_s$ satisfy \NEL\ and \AAU. Every agent $i$ has a fixed number $m_i$ of picks; write $\mathrm{pos}(i,t)$ for the turn of her $t$-th pick. At a unanimous profile the picks are made in the order of the common ranking, agent $i$'s $t$-th most preferred object being her $t$-th pick. We use three facts.

\emph{(a) Sizes.} Draft rules are non-wasteful, so by \NEL\ and \autoref{lem:sizes}, $m_i\in\{k,k+1\}$ with exactly $r$ agents of size $k+1$; write $A$ and $B$ for the agents of size $k$ and $k+1$.

\emph{(b) \AAU\ forces round structure inside each size class.} Let $C$ be $A$ or $B$ and $t$ such that every agent of $C$ has a $(t+1)$-th pick. At a unanimous profile, order $C$ by the position of the $t$-th pick; \AAU\ says this order is reversed at step $t+1$. Let $g$ be the last agent of $C$ at step $t$; then $g$ is first at step $t+1$, and for all $i,j\in C$, $\mathrm{pos}(i,t)\le\mathrm{pos}(g,t)<\mathrm{pos}(g,t+1)\le\mathrm{pos}(j,t+1)$. So all $t$-th picks of $C$ precede all $(t+1)$-th picks of $C$.

\emph{(c) \NEL\ forces $\mathrm{pos}(i,t)<\mathrm{pos}(j,t)$ for $i\in A$, $j\in B$, $t\le k$.} At a unanimous profile, remove $j$'s last pick: \NEL\ requires $x_i$ to dominate $j$'s first $k$ picks, which by \autoref{lem:sorted} is the stated position condition.

\emph{Sufficiency.} If $r=0$ there is a single class, and (b) gives full round structure; then $\Dr_s$ satisfies \EFO\ by the argument in \autoref{lem:draft}(b), which uses only round structure and $|m_i-m_j|\le1$. If $k=0$, \EFO\ is trivial. If $k=1$ and $r\ge1$, (c) puts every first pick of $A$ before every first pick of $B$, and (b) puts every first pick of $B$ before every second pick of $B$; hence all first picks precede all second picks, which is round structure, and \EFO\ follows as before.

\emph{Necessity.} Let $r\ge1$ and $k\ge2$, fix a draw $\pi$, and consider the draft rule $\Dr_{\tilde s}$ with
\[
\tilde s\;=\;A^1\,A^2\cdots A^k\;\;B^1\,B^2\cdots B^{k+1},
\]
in which the $A$-block completes all its rounds, snaking, before the $B$-block starts. It is a draft rule. \AAU\ holds because each block snakes within its own rounds and the two blocks have different sizes. \NEL\ holds because every pick of an $A$ agent precedes every pick of a $B$ agent, so for $i\in A$ and $j\in B$ \autoref{lem:draft}(a) gives $i_t\succ_i j_{t'}$ for all $t,t'$, and the injection used in the proof of \NEL\ in \autoref{lem:draft}(b) applies for every removed $y$. \EFO\ fails: at a unanimous profile every object of an $A$ agent $i$ is ranked above every object of a $B$ agent $j$, and $x_i\setminus\{y\}$ has $k-1\ge1$ elements for every $y$, none of which can be matched to an element of $x_j$ ranked at least as high.
\end{proof}

\autoref{thm:drafts} and \autoref{thm:strict}(b) remain true when non-wastefulness, neutrality, \EFO\ and \NEL\ are imposed only at unanimous profiles. For rules on $\cP^N$, when a condition among \NW, \EFO, \NEL, \NEU\ is required only at unanimous profiles, we write \NW$^u$, \EFO$^u$, \NEL$^u$, \NEU$^u$; for \NEU$^u$ the requirement is $f((\succ^u)^\sigma)=\sigma(f(\succ^u))$ for every unanimous profile and every bijection $\sigma$. \AAU\ is by definition a condition at unanimous profiles. The proof of \autoref{thm:drafts} in the text uses \NW\ and \NEU\ only at unanimous profiles, so it proves the stronger statement with \MM, \NW$^u$ and \NEU$^u$. We prove \autoref{thm:strict}(b) with \MM, \NW$^u$, \EFO$^u$, \NEL$^u$, \AAU\ and \NEU$^u$.

\begin{proof}[Proof of \autoref{thm:strict}]
(b) $\Dr^\pi$ satisfies the six conditions by \autoref{lem:draft}(b) and \autoref{lem:draftmm}, hence also the restricted ones. Conversely, let $f$ satisfy \MM, \NW$^u$, \EFO$^u$, \NEL$^u$, \AAU\ and \NEU$^u$. By \autoref{thm:drafts}, which uses \NW\ and \NEU\ only at unanimous profiles, the rule is $\Dr_s$ for some sequence $s$, and $s$ is read off any unanimous profile. It remains to show that the fairness conditions at unanimous profiles force $s=s(\pi)$ for some $\pi$. Step~1 reads a draw off a unanimous profile, and since the sequence $s$ is the same at every unanimous profile, that draw determines $s$.

If $k=0$, the sequence $s$ has length $T=r$, and by \NW$^u$, \EFO$^u$ and \autoref{lem:sizes} its entries are $r$ distinct agents; \NEL\ and \AAU\ impose nothing. Let $\pi$ list the $n-r$ agents who do not appear in $s$, in any order, followed by $s_1,\dots,s_r$. Then $s(\pi)=B^1=(s_1,\dots,s_r)$, so $f=\Dr^\pi$; if also $T=0$, every rule is the empty rule and every $\pi$ works. Assume $k\ge1$ from now on, so that every agent has a first pick.

\smallskip\noindent\emph{Step 1: unanimous profiles.}
Fix a unanimous profile with common order $\succ^u$ given by $z_1\succ^u z_2\succ^u\cdots\succ^u z_T$, and let $x=f(\succ^u)$. All dominance comparisons in this step are with respect to $\succ^u$, so \autoref{lem:sorted} applies.

\emph{(a) Sizes.} By \NW$^u$, \EFO$^u$ and \autoref{lem:sizes}, applied at $\succ^u$, exactly $r$ agents receive $k+1$ objects. Call them $B$ and the others $A$.

\emph{(b) Round structure.} For $i\neq j$, by \autoref{lem:sorted} the sorted list of $x_j\setminus\{y\}$ is pointwise $\succsim^u$ the sorted list of $x_j\setminus\{x_j(1)\}$ for every $y\in x_j$; hence \EFO\ for the pair $(i,j)$ is equivalent to $x_i(\ell)\succ^u x_j(\ell+1)$ for all $\ell\le|x_j|-1$. Let $R_t=\{x_i(t): i\in N,\ |x_i|\ge t\}$. Every element of $R_t$ is ranked above every element of $R_{t+1}$: for $x_i(t)$ and $x_j(t+1)$ with $i\neq j$ this is \EFO\ with $\ell=t$, and for $i=j$ it holds by definition. The sets $R_1,\dots,R_{k+1}$ partition $O$ with $|R_t|=n$ for $t\le k$ and $|R_{k+1}|=r$, so
\[
R_t=\{z_{(t-1)n+1},\dots,z_{tn}\}\ (t\le k),\qquad R_{k+1}=\{z_{kn+1},\dots,z_T\}.
\]

\emph{(c) Blocks.} Let $i\in A$ and $j\in B$. \NEL\ with $y=x_j(k+1)$ gives $x_i\dom{}\{x_j(1),\dots,x_j(k)\}$, i.e.\ $x_i(t)\succ^u x_j(t)$ for all $t\le k$ by \autoref{lem:sorted}. Hence inside each $R_t$, $t\le k$, the objects of $A$ agents precede the objects of $B$ agents.

\emph{(d) Reversal.} For a nonempty block $C\in\{A,B\}$ with common bundle size $q_C$, and $1\le t\le q_C$, let $\lhd_t$ be the linear order on $C$ defined by $i\lhd_t j\iff x_i(t)\succ^u x_j(t)$. For $t<q_C$, \AAU\ and distinctness of objects imply that $i\lhd_t j$ gives $j\lhd_{t+1}i$ for all pairs, so $\lhd_{t+1}$ is the reverse of $\lhd_t$.

Define $\pi_u(i)$ as the position of $x_i(1)$ in $z_1,\dots,z_n$. By (c), $A=A(\pi_u)$ and $B=B(\pi_u)$, and $\lhd_1$ on each block is the restriction of $\pi_u$. By (b)--(d), the objects of $R_t$ are assigned, in the order of $\succ^u$, to the $A$ agents in the order $\lhd_t$ and then to the $B$ agents in the order $\lhd_t$, where $\lhd_t$ is the restriction of $\pi_u$ reversed $t-1$ times; and $R_{k+1}$ is assigned to the $B$ agents in the order $\lhd_{k+1}$. This is exactly round $t$ of $s(\pi_u)$. Since $\Dr^{\pi_u}$ at $\succ^u$ picks $z_1,z_2,\dots$ in order, we conclude $x=\Dr^{\pi_u}(\succ^u)$.

\smallskip\noindent\emph{Conclusion.} At every unanimous profile, $f=\Dr_s$ gives the $m$-th ranked object to the $m$-th picker of $s$. Step~1 at a single unanimous profile $\succ^u$ therefore gives $s=s(\pi_u)$, and $f=\Dr_s=\Dr^{\pi_u}$.

(a) If $s=s(\pi)$, then $\Dr_s=\Dr^\pi$ satisfies \EFO, \NEL\ and \AAU\ by \autoref{lem:draft}(b). Conversely, let $\Dr_s$ satisfy them. Every draft rule is non-wasteful and neutral and satisfies \MM\ (\autoref{thm:drafts}), so by part (b), $\Dr_s=\Dr^\pi$ for some draw $\pi$. A draft rule gives the $m$-th ranked object to its $m$-th picker at every unanimous profile, so $s=s(\pi)$.
\end{proof}

\begin{proof}[Proof of \autoref{thm:cf}]
(a) $L_s(\succsim)\neq\emptyset$ by \autoref{lem:fl}(a), and every allocation in it is non-wasteful by \autoref{lem:fl}(c). PD-efficiency: let $x\in L_s(\succsim)$ with injection $\varphi$, and suppose $y$ PD-dominates $x$. The idea is to line up $y$ with $x$ turn by turn, take the earliest turn at which $y$ gives the picker an object she strictly prefers, and show that the forward-looking draft left that object available, a contradiction. Since $x$ exhausts $O$ and $|y_i|\ge|x_i|$ for all $i$ (dominance requires an injection), $|y_i|=|x_i|$ for all $i$. For each agent $i$ with turns $m_1<\cdots<m_q$, list $y_i$ in weakly decreasing $\succsim_i$-order and assign its $\ell$-th element to turn $m_\ell$; this defines an injection $\psi:\{1,\dots,T\}\to O$. By \autoref{lem:sorted} and \autoref{lem:fl}(a), $\psi(m)\succsim_{s_m}\varphi(m)$ for every $m$, and for the agent $j$ with $y_j\sdom{j}x_j$ there is a turn with strict preference (otherwise $x_j\dom{j}y_j$). Let $m^*$ be the earliest turn with $\psi(m^*)\succ_{s_{m^*}}\varphi(m^*)$. For $m<m^*$ we have $\psi(m)\sim_{s_m}\varphi(m)$, i.e.\ $\psi(m)\in K_m$, and \autoref{lem:fl}(b) with $\bar m=m^*$ gives $\psi(m)\in C_m$ for all $m<m^*$. So $\psi|_{\{1,\dots,m^*-1\}}$ is an $m^*$-partial assignment avoiding $\psi(m^*)$, whence $\psi(m^*)\in F_{m^*}$. But $\varphi(m^*)$ is $\succsim_{s_{m^*}}$-maximal in $F_{m^*}$, contradicting $\psi(m^*)\succ_{s_{m^*}}\varphi(m^*)$. On strict profiles, where $L_s(\succ)=\{\Dr_s(\succ)\}$ by \autoref{lem:fl}(c), it follows that every draft rule is PD-efficient.

(b) and (c) These are \autoref{lem:fl}(c). For $s=s(\pi)$, $\Dr^\pi$ satisfies \EFO\ and \NEL\ by \autoref{lem:draft}(b), so (b) applies to $L^\pi$, and the bundle sizes hold by construction of $s(\pi)$.
\end{proof}

\begin{proof}[Proof of \autoref{prop:nelsp}]
(a) Let $\Dr_s$ satisfy \NEL\ and \WSP. By \autoref{lem:manip} every agent's picks are consecutive. Draft rules are non-wasteful, so by \NEL\ and \autoref{lem:sizes}, $n-r$ agents have $k$ picks and $r$ have $k+1$; if $k=0$, the agents with zero picks can be placed first without changing the rule. If $k\ge1$, fact (c) in the proof of \autoref{cor:ef1} shows that, for $i$ with $k$ picks and $j$ with $k+1$, $i$'s $t$-th pick precedes $j$'s $t$-th pick for every $t\le k$. In this case, with consecutive picks, this holds for all $t$ if and only if $i$'s run of picks starts before $j$'s, that is, precedes it entirely. Conversely, in a quota serial dictatorship with the $k$-agents first, every pick of a $k$-agent $i$ precedes every pick of a $(k+1)$-agent $j$, so by \autoref{lem:draft}(a) $i$ prefers each of her objects to each of $j$'s; any bijection from $x_j\setminus\{y\}$ onto $x_i$ then witnesses $x_i\dom{i}x_j\setminus\{y\}$ for every $y$, which is \NEL. Strategy-proofness holds because each agent, choosing consecutively, receives her favorite remaining objects up to her quota, which dominate every other bundle of that size she could obtain; and as a draft rule a quota serial dictatorship also satisfies \NW, \NEU, \EFF\ and \MM, which gives the last claim.

(b) Suppose that at least two agents receive more than one object, and let $\Dr_s$ satisfy \EFO. Draft rules are non-wasteful, so by \autoref{lem:sizes} the bundle sizes are $k$ and $k+1$, and by Step~1(b) of the proof of \autoref{thm:strict}(b), applied at a unanimous profile, every agent's $t$-th pick comes before every agent's $(t+1)$-th pick. The hypothesis gives two agents $i$ and $j$ who pick at least twice: all agents if $k\ge2$, and the $r\ge2$ agents with two objects if $k=1$. If $i$'s first pick comes before $j$'s, then $j$'s first pick lies strictly between $i$'s first and second picks, and $i$ has no turn in between. By \autoref{lem:manip}, $\Dr_s$ is not weakly strategy-proof.
\end{proof}

\begin{proof}[Proof of \autoref{prop:imposs}]
The idea is as follows. At unanimous profiles, non-wastefulness and neutrality reduce the rule to a picking sequence, and \EFO\ makes that sequence proceed in rounds, so it remains to find a profile at which some agent can profitably misreport. We let one agent, $a$, report a reference order and the other, $b$, report a second order, chosen so that at the unanimous profile with the second order, $a$ receives her most preferred objects under the reference order. Since $a$ could obtain these objects by reporting the second order herself, weak strategy-proofness restricts her bundle at the mixed profile (\autoref{lem:topset}), and with it $b$'s. The second order is built so that $b$, by reporting the reference order instead, moves to the unanimous profile with that order and obtains a bundle she strictly prefers object by object. The three cases below differ only in how the second order is built.

Suppose $f$ satisfies the four conditions. Since $n=2$, $r\in\{0,1\}$, and by \NW, \EFO\ and \autoref{lem:sizes} one agent receives $k$ objects and the other $k+r$ at every profile.

\smallskip\noindent\emph{Pattern.} By the proof of \autoref{thm:drafts}, which uses only \NW\ and \NEU\ at unanimous profiles, there is a sequence $s\in N^T$ such that at every unanimous profile with common order $z_1\succ\cdots\succ z_T$ agent $s_m$ receives $z_m$. Write $\mathrm{Pos}_i=\{m:s_m=i\}$ for the ranks agent $i$ receives. By Step~1(b) of the proof of \autoref{thm:strict}(b), each pair of ranks $\{2t-1,2t\}$, $t=1,\dots,k$, contains one rank of each agent, and if $r=1$ the rank $T$ belongs to the agent who receives $k+1$ objects. Let $\succ^0$ be the order $z_1\succ z_2\succ\cdots\succ z_T$, let $Q=(\succ^0,\succ^0)$ and $x=f(Q)$, so that $x_i=\{z_m:m\in\mathrm{Pos}_i\}$.

\smallskip\noindent\emph{Scheme.} In each case below we choose an agent $a$, call the other agent $b$, and construct an order $\succ^2$, specified by the object at each rank, under which the ranks in $\mathrm{Pos}_a$ are occupied by the set $S$ of the $|\mathrm{Pos}_a|$ most preferred objects under $\succ^0$. At $Q'=(\succ^2,\succ^2)$ agent $a$ therefore receives $S$. Let $P$ be the profile at which $a$ reports $\succ^0$ and $b$ reports $\succ^2$. At $P$, agent $a$ can report $\succ^2$, which produces $Q'$ and gives her $S$, the set of her $|S|$ most preferred objects, so \autoref{lem:topset} restricts $f_a(P)$. Agent $b$ can report $\succ^0$, which produces $Q$ and gives her $x_b$. In each case $x_b\sdom{}f_b(P)$ under $\succ^2$, so $b$ has a profitable deviation at $P$, contradicting \WSP.

\smallskip\noindent\emph{Counting claim.} Let $u_1<\cdots<u_k$ and $v_1<\cdots<v_k$ be distinct ranks with $u_j<v_\ell$ whenever $j<\ell$, and let $1\le m\le k-1$. If, under some order, $X$ occupies the ranks $\{u_1,\dots,u_m\}\cup\{v_1,\dots,v_{k-m}\}$ and $Y$ the ranks $\{v_1,\dots,v_k\}$, then $X\sdom{}Y$. Match the objects of $Y$ at ranks $v_1,\dots,v_{k-m}$ to themselves in $X$. For $j=1,\dots,m$, match the object at rank $v_{k-m+j}$ to the object at rank $u_j$. These images are distinct and exhaust $X$. Since $k-m\ge1$, we have $j<k-m+j$, and hence $u_j<v_{k-m+j}$ by the assumed cross-rank inequality. The map therefore witnesses $X\dom{}Y$. The dominance is strict because $X$ and $Y$ are distinct sets of the same size under a strict order.

\smallskip\noindent\emph{Case $r=0$.} Let $a=s_1$, and write $\mathrm{Pos}_a=\{p_1<\cdots<p_k\}$ and $\mathrm{Pos}_b=\{q_1<\cdots<q_k\}$. Then $p_t,q_t\in\{2t-1,2t\}$, $p_1=1$, $q_1=2$, and $p_j\le2j\le2\ell-2<q_\ell$ for $j<\ell$. Here $S=\{z_1,\dots,z_k\}$; let $m=|x_b\cap S|$. Since $k\ge2$, $z_2\in x_b\cap S$, so $m\ge1$; since $z_1\in x_a\cap S$, $m\le k-1$. Let $\succ^2$ place the objects of $x_b\cap S$ at the ranks $p_1,\dots,p_m$, those of $x_a\cap S$ at $p_{m+1},\dots,p_k$, those of $x_b\setminus S$ at $q_1,\dots,q_{k-m}$, and those of $x_a\setminus S$ at $q_{k-m+1},\dots,q_k$, in any order within each group (the group sizes are $m$, $k-m$, $k-m$ and $m$). By \autoref{lem:topset}(a), $f_a(P)=S$, so $f_b(P)=O\setminus S$, which occupies the ranks $q_1,\dots,q_k$ under $\succ^2$, while $x_b$ occupies $\{p_1,\dots,p_m\}\cup\{q_1,\dots,q_{k-m}\}$. The counting claim with $u=p$ and $v=q$ gives $x_b\sdom{}f_b(P)$ under $\succ^2$.

\smallskip\noindent\emph{Case $r=1$ and $k\ge3$.} Let $a$ be the agent who receives $k+1$ objects at unanimous profiles, and write $\mathrm{Pos}_a=\{q_1<\cdots<q_{k+1}\}$ and $\mathrm{Pos}_b=\{p_1<\cdots<p_k\}$. Then $p_t,q_t\in\{2t-1,2t\}$ for $t\le k$, $q_{k+1}=T$, and $q_j\le2j\le2\ell-2<p_\ell$ for $j<\ell\le k$. Here $S=\{z_1,\dots,z_{k+1}\}$; let $m=|x_b\cap S|$. Since $p_1\le2$, $m\ge1$; since $p_k\ge2k-1>k+1$ for $k\ge3$, $m\le k-1$. Let $\succ^2$ place the objects of $x_b\cap S$ at the ranks $q_1,\dots,q_m$, those of $x_a\cap S$ at $q_{m+1},\dots,q_{k+1}$, those of $x_b\setminus S$ at $p_1,\dots,p_{k-m}$, and those of $x_a\setminus S$ at $p_{k-m+1},\dots,p_k$ (the group sizes are $m$, $k+1-m$, $k-m$ and $m$). By \autoref{lem:topset}(a), $f_a(P)=S$, so $f_b(P)=O\setminus S$, which occupies the ranks $p_1,\dots,p_k$, while $x_b$ occupies $\{q_1,\dots,q_m\}\cup\{p_1,\dots,p_{k-m}\}$. The counting claim with $u=q$ and $v=p$ gives $x_b\sdom{}f_b(P)$ under $\succ^2$.

\smallskip\noindent\emph{Case $r=1$ and $k=2$.} Let $a$ be the agent who receives two objects at unanimous profiles, and write $\mathrm{Pos}_a=\{p_1,p_2\}$ and $\mathrm{Pos}_b=\{q_1,q_2,5\}$ with $p_1,q_1\in\{1,2\}$ and $p_2,q_2\in\{3,4\}$, so that $x_a=\{z_{p_1},z_{p_2}\}$ and $x_b=\{z_{q_1},z_{q_2},z_5\}$. Here $S=\{z_1,z_2\}=\{z_{p_1},z_{q_1}\}$. Let $\succ^2$ place $z_{q_1}$ at rank $p_1$, $z_{p_1}$ at rank $p_2$, $z_5$ at rank $q_1$, and $z_3$ and $z_4$ at the ranks $q_2$ and $5$ in either order. Under $\succ^2$ the two most preferred objects, at ranks $1$ and $2$, are $z_{q_1}$ and $z_5$, both in $x_b$. By \autoref{lem:topset}(b), either $f_a(P)=S$ or $f_a(P)$ has three objects. In the first case $f_b(P)=\{z_3,z_4,z_5\}$, with ranks $q_1<q_2<5$ under $\succ^2$, while the ranks of $x_b$ are $1$, $2$ and a rank at most $5$; since $q_2\ge3$, $x_b\dom{}f_b(P)$ by \autoref{lem:sorted}, strictly because the two sets have the same size and $z_{q_1}\in x_b\setminus f_b(P)$. In the second case $f_b(P)$ has two objects, and $x_b$, which contains the two most preferred objects under $\succ^2$, dominates it, strictly because a set cannot dominate a larger set.
\end{proof}

\begin{proof}[Proof of \autoref{prop:nom}]
Fix a picking sequence $s$, let $f=\Dr_s$, and fix an agent $i$, her true order $\succ_i$ and an additive utility $v_i$ with positive values consistent with $\succ_i$. For a report $\succ'_i$ write $\underline v_i(\succ'_i)=\min_{\succ_{-i}}v_i(f_i(\succ'_i,\succ_{-i}))$ and $\overline v_i(\succ'_i)=\max_{\succ_{-i}}v_i(f_i(\succ'_i,\succ_{-i}))$; we show that $\underline v_i(\succ'_i)\le\underline v_i(\succ_i)$ and $\overline v_i(\succ'_i)\le\overline v_i(\succ_i)$ for every $\succ'_i$. Let $i$ have turns $m_1<\cdots<m_q$ in $s$ (if $q=0$ there is nothing to prove), let $\mathrm{Top}_h$ be the set of the $h$ most preferred objects under $\succ_i$, and let $W$ be the bundle of the $m_1$-th, \dots, $m_q$-th most preferred objects under $\succ_i$. Truth-telling guarantees the agent at least $W$, while for every report there is a profile at which she receives a bundle that $W$ dominates; and in the best case truth-telling gives her her $q$ most preferred objects.

\emph{Worst case.} At turn $m_\ell$ at most $m_\ell-1$ objects have been taken, so under truth-telling $i$ weakly prefers her $\ell$-th pick to her $m_\ell$-th most preferred object, whatever the others report; mapping the $m_\ell$-th most preferred object to the $\ell$-th pick shows that $i$'s bundle dominates $W$, so $\underline v_i(\succ_i)\ge v_i(W)$. Now let every other agent report $\succ_i$, so that every pick by another agent takes the remaining object that $\succ_i$ ranks first, and let $y$ be the bundle $i$ receives when she reports $\succ'_i$. We claim that $|y\cap \mathrm{Top}_h|\le\nu(h)$ for every $h$, where $\nu(h)=|\{\ell:m_\ell\le h\}|$ is the number of $i$'s turns among the first $h$. If every object of $y\cap \mathrm{Top}_h$ is picked at one of $i$'s first $\nu(h)$ turns, this is immediate. Otherwise let $m_t>h$ be the last turn at which $i$ picks an object $z\in \mathrm{Top}_h$, and let $c=|y\cap \mathrm{Top}_h|$. Each of the $m_t-t$ picks by other agents before turn $m_t$ was made while $z$ was available, so it took an object ranked above $z$, which lies in $\mathrm{Top}_h$; together with the $c$ objects of $y\cap \mathrm{Top}_h$, all picked by turn $m_t$, this gives $m_t-t+c\le h$. Since $m_{\nu(h)+1}<\cdots<m_t$ are $t-\nu(h)$ distinct integers larger than $h$, $m_t\ge h+t-\nu(h)$, and so $c\le\nu(h)$. By the claim, the $\ell$-th most preferred object of $y$ is not among the $m_\ell-1$ most preferred objects, since otherwise $|y\cap \mathrm{Top}_{m_\ell-1}|\ge\ell>\nu(m_\ell-1)$. Hence $W$ dominates $y$ by \autoref{lem:sorted}, and $\underline v_i(\succ'_i)\le v_i(y)\le v_i(W)\le\underline v_i(\succ_i)$.

\emph{Best case.} Let every other agent rank the objects outside $\mathrm{Top}_q$ above those in $\mathrm{Top}_q$. Under truth-telling, by induction on the turns, the other agents take only objects outside $\mathrm{Top}_q$ and $i$ takes only objects of $\mathrm{Top}_q$: at each of the others' turns fewer than $T-q$ objects outside $\mathrm{Top}_q$ have been taken, and at each of $i$'s turns fewer than $q$ objects of $\mathrm{Top}_q$ have been taken. So $i$ receives $\mathrm{Top}_q$, which dominates every bundle of $q$ objects, and $\overline v_i(\succ'_i)\le v_i(\mathrm{Top}_q)=\overline v_i(\succ_i)$.
\end{proof}

\begin{proof}[Proof of \autoref{prop:spnb}]
(a) Let $x=f(\succ)$, fix an agent $i$, and let $\succ'_i$ satisfy, for every $y\in x_i$, $\{z:y\succ_iz\}\subseteq\{z:y\succ'_iz\}$. Write $x'=f(\succ'_i,\succ_{-i})$. By \SP\ at $\succ$, there is an injection $\mu:x'_i\to x_i$ with $\mu(z)\succsim_iz$ for every $z\in x'_i$. If $\mu(z)\neq z$, then $\mu(z)\succ_iz$ because $\succ_i$ is a linear order, and since $\mu(z)\in x_i$ the hypothesis gives $\mu(z)\succ'_iz$. Hence $x_i\succsim^{\PD}x'_i$ under $\succ'_i$. By \SP\ at $(\succ'_i,\succ_{-i})$, $x'_i\succsim^{\PD}x_i$ under $\succ'_i$. The two bundles therefore have the same size, and by \autoref{lem:sorted} their $\ell$-th most preferred elements under the linear order $\succ'_i$ coincide for every $\ell$, so $x'_i=x_i$. By \NB, $f(\succ'_i,\succ_{-i})=x$. The hypothesis of \MM\ is a condition on each agent's own ranking relative to her own bundle at the fixed allocation $x$, so changing the agents' rankings one agent at a time and applying this argument each time gives $f(\succ')=x$.

(b) Let $f$ satisfy \NW, \NEU, \SP\ and \NB. By (a) it satisfies \MM, so $f=\Dr_s$ for some picking sequence $s$ by \autoref{thm:drafts}. Strategy-proofness implies weak strategy-proofness, since a deviation bundle that strictly dominates the truthful bundle cannot be dominated by it. If some agent had two turns in $s$ with no turn of her own and at least one turn of another agent strictly between them, \autoref{lem:manip} would contradict weak strategy-proofness. Hence every agent's picks are consecutive, and $\Dr_s$ is a quota serial dictatorship. Conversely, let $f$ be a quota serial dictatorship. It is a draft rule, so it satisfies \NW\ and \NEU. The agents who choose before $i$ do not depend on $i$'s report, and $i$ receives her favorite remaining objects up to her quota, which dominate every other bundle of that size from the same set, which is all she can obtain by misreporting; so $f$ is strategy-proof. The agents who choose after $i$ face the objects left by those before them and by $i$, which depend on $i$'s report only through her bundle; so $f$ is nonbossy.
\end{proof}

\begin{proof}[Proof of \autoref{cor:nb}]
By \autoref{prop:spnb}(b), $f$ is a quota serial dictatorship. Being strategy-proof, it is weakly strategy-proof, so by \autoref{prop:nelsp}(b) it violates \EFO\ whenever at least two agents receive more than one object.
\end{proof}

\end{document}